\documentclass[12pt,aps,prd,amsmath,amssymb, showpacs]{revtex4}
\usepackage{graphicx}% Include figure files
\usepackage{bm}% bold math

\newcommand{\beq}{\begin{equation}}
\newcommand{\beqn}{\begin{eqnarray}}
\newcommand{\eeq}{\end{equation}}
\newcommand{\eeqn}{\end{eqnarray}}

\begin{document}
\title{Density Perturbations in Hybrid Inflation 
Using a Free Field Theory Time-Delay Approach}
\author{Alan H. Guth 
and Evangelos I. Sfakianakis}
\email{guth@ctp.mit.edu; esfaki@mit.edu}
\affiliation{Center for Theoretical Physics, Laboratory for Nuclear Science and Department of Physics, 
\\ Massachusetts Institute of Technology, Cambridge, MA 02139, USA}
\date{\today}

\begin{abstract}

We introduce a new method for calculating density perturbations
in hybrid inflation which avoids treating the fluctuations of the
``waterfall'' field as if they were small perturbations about a
classical trajectory.  We quantize only the waterfall field,
treating it as a free quantum field with a time-dependent $m^2$,
which evolves from positive values to tachyonic values.  Although
this potential has no minimum, we think it captures the important
dynamics that occurs as $m^2$ goes through zero, at which time a
large spike in the density perturbations is generated.  We assume
that the time-delay formalism provides an accurate approximation
to the density perturbations, and proceed to calculate the power
spectrum of the time delay fluctuations.  While the evolution of
the field is linear, the time delay is a nonlinear function to
which all modes contribute.  Using the Gaussian probability
distribution of the mode amplitudes, we express the time-delay
power spectrum as an integral which can be carried out
numerically.  We use this method to calculate numerically the
spectrum of density perturbations created in hybrid inflation
models for a wide range of parameters.  A characteristic of the
spectrum is the appearance of a spike at small length scales,
which can be used to relate the model parameters to observational
data. It is conceivable that this spike could seed the formation
of black holes that can evolve to become the supermassive black
holes found at the centers of galaxies. 

\end{abstract}

\pacs{Preprint MIT-CTP 4415}

\maketitle

\section{Introduction}

Inflation
\cite{Guth:1980zm}
remains the leading paradigm for the very early universe. It
naturally solves the cosmological flatness and horizon problems
and is consistent with high precision measurements of the cosmic
microwave background radiation \cite {inflation,WMAP}. Numerous
models of inflation have been proposed, each adding features to
the predictions of a scale invariant spectrum derived from
single-field slow-roll inflation. Their motivation can be either
some particle physics ideas coming from the standard model
\cite{higgs} or supersymmetric theories \cite{sol1,sol2}, the
need to explain some observation such as glitches in the CMB or
supermassive black holes in galactic centers, or simply the
extension of a theorist's toolbox in anticipation of the next set
of high precision data, such as the upcoming Planck satellite
measurements.

Hybrid inflation was first proposed by A. Linde \cite{linde} and
the name was chosen because this class of models can be thought
of as being a hybrid between chaotic inflation and inflation in a
theory with spontaneous symmetry breaking. The simplest hybrid
inflation model requires two fields that we will call the timer
and waterfall fields. The timer field corresponds the usual slow
rolling field and is responsible for the scale invariant spectrum
of perturbations observed in the CMB. The waterfall field is
confined to its origin by the interaction with the timer field,
giving a large constant contribution to the potential, which is
also the main contribution to the energy density and hence the
Hubble parameter.  The potential governing the waterfall field
changes as the timer field evolves, and at some point the minimum
of the potential turns into a local maximum, and the waterfall
field rolls down its tachyonic potential to its new minimum,
where inflation ends.  A characteristic feature of the density
perturbation spectrum of hybrid inflation is the appearance of a
large spike generated at the time when the waterfall potential
turns tachyonic.  The spike is generically at small length
scales, and can potentially seed primordial black holes \cite
{lindeBH}.  Primordial black hole formation and evolution has
been studied in the past
\cite{carr1,carr2,carr3,carr4}, but whether these black holes
grow to become the supermassive black holes currently found in
galactic centers is an open and intriguing possibility that we
will address in a future publication.

Usual inflationary perturbation theory is based on the study of
quantum fluctuations around a classical trajectory in field
space.  However, in a purely classical formulation the waterfall
field of hybrid inflation would remain forever at the origin,
even after the waterfall transition, due to symmetry. It is
quantum fluctuations that destabilize it and lead to the end of
inflation, so in a sense the classical trajectory has a quantum
origin.  Numerous papers have used various analytical approaches
or numerical simulations to overcome this difficulty and
approximate the spectrum of density perturbations
\cite{sol1,sol2, lyth1, lyth2, lyth3, stoch, sasaki,
kodama,wands,kristin,son}. 

The method we use here has evolved from the early work in Kristin
Burgess' thesis \cite{kristin}, in which she studied a free-field
model of the waterfall field in one space dimension, focusing on
the time delay of the scalar field as a measure of perturbations. 
As in the model considered here, the waterfall field was
described by a Lagrangian with a time-dependent $m^2$, caused by
the interaction with the timer field.  $m^2$ evolved from
positive values at early times to negative (tachyonic) values at
late times.  Such models are unnatural, since the potential is
not bounded from below, but they nonetheless appear to be useful
toy models, since the dynamics that generate the spike in the
fluctuation spectrum occur during the transition from positive to
negative $m^2$.  The evolution in the bottomless potential is
realistic enough to give a well-defined time delay.  Burgess
studied the evolution of the waterfall field by means of a
numerical simulation on a spatial lattice, using 262,000 points,
calculating the power spectrum of the time delay by Monte Carlo
methods.  The method was slow, but for one choice of parameters
she accumulated 5000 runs, giving a very reliable graph of the
time-delay spectrum for this model.  This line of research was
pursued further in the thesis of Nguyen Thanh Son \cite{son}, who
repeated Burgess' numerical simulations with a new code (with
excellent agreement).  More importantly, Son and one of us (AHG),
with some crucial input from private communication with Larry
Guth, developed a method to short-circuit the Monte Carlo
calculation.  Instead of determining the power spectrum by
repeated random trials, it was possible to express the
expectation value for the random trials as an explicit expression
involving integrals over mode functions, which could be evaluated
numerically.  The speed and numerical precision were dramatically
improved.  While Son's work was still limited to one spatial
dimension, the possibility of extending it to three spatial
dimensions was now a very realistic goal.  In this paper we
extend the calculation of the time-delay power spectrum in
free-field models of hybrid inflation to three spatial
dimensions, calculating the spectrum for a wide range of model
parameters.

In Section II we define the free-field model for the timer and
    waterfall fields that we will use to calculate fluctuations. We
    set up the equations of motion, define the notation of the mode
    expansion, and discuss the behavior of the mode functions.  We
    make contact with a class of supersymmetric models that support
    hybrid inflation in Section III, presenting the form of their
    potential and the range of parameters that they allow.  Section
    IV gives a brief summary of the time delay formalism, and
    presents an approximation for calculating perturbations,
    developed earlier by Randall, Solja\v{c}i\'{c}, and Guth. In
    Section V we develop a new method for calculating density
    perturbations in hybrid inflation that avoids any need to
    consider small fluctuations about a classical solution. Instead
    we show how the time delay power spectrum can be calculated
    essentially exactly in the context of the free field theory
    description.  The result is given in the form of an integral over
    the modes which makes use of their known Gaussian probability
    distribution. In section VI we present an extensive set of
    numerical results over the parameter space of our model, where we
    are able to isolate the main factors that influence the density
    perturbation spectrum. In the limit of a light timer field, all
    quantities of interest are determined by the product of the timer
    and waterfall masses. We examine the models discussed in Section
    II as examples of realistic versions of hybrid inflation, and
    provide graphs showing the predictions of these models. 
    Concluding remarks and directions of future work follow in
    Section VII. 
\section{Model}

\subsection{Field set-up}

Our first assumption is related to the expansion rate. We
consider the metric to be exactly De-Sitter, even though this is
only approximately correct. However, it changes only weakly
during the slow roll inflation era, and we will terminate our
calculation once the approximation loses its validity. Defining
the Hubble constant during inflation as $H$, the scale factor is
written as
\beq
a(t) = e^{Ht}
\eeq
The model consists of two scalar fields. The "waterfall" field
$\phi$ with lagrangian
\begin{equation}  L_\phi=e^{3Ht} \left [ |\dot \phi|^2 -  e^{-2Ht}|\nabla \phi|^2 - m_\phi ^2 (t) |\phi|^2 \right ] 
\label{lphi}
\end{equation}
The usual $1/2$ factors can be restored, if one writes $\phi =
{1\over \sqrt 2} (\phi_1+i\phi_2)$ where $\phi_1$ and $\phi_2$
are real scalar fields. The waterfall field must be complex,
otherwise it will create domain walls as it rolls down from its
initial value. The time-dependent mass of the $\phi$ field is
controlled by a real scalar field, subsequently called the
"timing" field. The important property of the squared mass of
$\phi$ is that it has to be positive initially and as $\psi$
evolves become negative. A general form is the following
\begin{equation} 
m_\phi ^2 (t) = -m_0^2 \left [ 1-\left ( { \psi(t)\over \psi _c }\right ) ^r \right ] 
\end{equation}
We will choose $r=4$ for most of our simulations.  The lagrangian
of the timing field is
\begin{equation} 
L_\psi=e^{3Ht} \left [ {1\over 2} \dot \psi^2 - {1\over 2} e^{-2Ht} (\nabla \psi)^2 -{1\over 2} m_ \psi ^2  \psi ^2 \right ]
\end{equation}
The Lagrangians define the system up to an additive constant
$V_0$ in the potential, which is taken to be large enough, so
that the variations in $H$ are negligible during the era of
interest.
We neglected the interaction term from the Lagrangian of the
timing field. This means that there is no back-reaction from the
waterfall to the timing field. Physically this is a reasonable
approximation before the waterfall transition, as well as
afterwards, for as long as the waterfall field remains close to
the origin. Mathematically, neglecting this term makes the
equation of motion for the timing field de-coupled and in our
quadratic approximation analytically solvable. 
\newline Furthermore we do not examine perturbations arising from
quantum fluctuations of the timing field. Before the waterfall
transition they will give the nearly scale invariant spectrum
that can be matched to the CMB observations. Apart from making
sure that the long wavelength tail of the waterfall field
perturbations does not contradict WMAP data, we will not consider
these scales.  After the waterfall transition the timer field
perturbations will continue to be of the order of $10^{-5}$,
hence they will be subdominant to the perturbations of the
waterfall field by a few orders of magnitude, as we will see.
The equations of motion are
\begin{eqnarray}
\ddot \phi + 3H\dot \phi -e^{-2Ht} \nabla^2 \phi = -m_\phi ^2 (t) \phi 
\label{eqn:phieq}
\\
\ddot \psi + 3H \dot \psi -e^{-2Ht} \nabla^2 \psi = -m_ \psi ^2  \psi 
\end{eqnarray}
If we take the timing field to be spatially homogenous, we get
 \beq
 \psi(t) = \psi_c e^{pt} ~,~ p= H\left (-{3\over 2} \pm\sqrt { {9\over 4} - {m_{\psi}^2 \over H^2}} \right )
 \eeq
The value of the constant of integration was chosen so that
$\psi(t)=\psi_c$ and $m_\phi ^2(t) = 0$ at $t=0$.
Both roots are negative, but the long time behavior is dominated
by the larger of the two roots, which is
\beq
p= -H\left ({3\over 2} -\sqrt { {9\over 4} - {m_\psi^2 \over
H^2}} \right )
\eeq
We will always choose ${m_\psi \over H} < 3/2$ and not consider
the case of a complex root. In fact, hybrid inflation models
usually require the mass of the timing field to be well below the
Hubble parameter, as in \cite {sol1} and \cite {sol2}.
We choose to measure time in number of e-folds, hence we use
$N=Ht$. We rescale the masses similarly as $\mu_\psi = m_\psi /H$
and $\mu_\phi = m_0/H$. Furthermore the finite box size that we
will use in our simulations is measured in units of ${1\over H}$
and the field magnitude in units of $H$.
We also define
\beq
\tilde \mu _\psi ^2 = -{rp\over H} = r\left ({3\over 2} -\sqrt { {9\over 4} - {m_\psi^2 \over H^2}} \right )
\eeq
For a light timer field the reduced mass $\tilde \mu_\psi$ is
proportional to the actual timer mass, $\tilde \mu_\psi = \sqrt
{r \over 3} \left ( {m_\psi \over H} \right ) = \sqrt {r \over 3}
\mu_\psi $.

\subsection{Fast Transition}
Let's consider the speed of the transition. The transition
happens at $m_\phi ^2 =0$. In order to quantify the speed of the
transition, we will use the basic scale of our system, the Hubble
scale. We will consider the transition duration to be the period
for which $|m_\phi| \le H$, meaning that the mass term in the
equation of motion of the waterfall field is negligible.
Assuming that $\tilde \mu_\phi > 1$ we get
\beq
\pm 1 = \mu_\phi^2 \left ( 1- e^{-\tilde  \mu_\psi^2 N}\right ) \Rightarrow    \Delta N ={1\over \tilde \mu_\psi^2} \log\left ({\mu_\phi ^2 +1 \over \mu_\phi ^2 - 1}\right ) 
\eeq
In the limit of $\tilde \mu_\phi \gg 1$
\beq 
 \Delta N ={2 \over (\tilde \mu_\psi \mu_\phi ) ^2}
\eeq
Another measure of the speed of the transition is given by
derivative of the waterfall field mass at $N=0$.
\beq
{1\over H^2} {dm_\phi^2(N) \over dN} |_{N=0}= (\tilde \mu_\psi
\mu_\phi ) ^2 \Rightarrow \Delta N \sim {1 \over (\tilde \mu_\psi
\mu_\phi ) ^2}
\eeq
This shows that as long as the product $\mu_\phi \mu_\psi$ is
somewhat larger than unity, the duration of the transition will
be less than a Hubble time, meaning that the transition is fast!

\subsection{Mode expansion}

For purposes of our numerical calculations, we think of the
universe as a finite box with periodic boundary conditions and a
discrete spatial lattice. We choose the lattice to be cubic with
length $b$ and $Q^3$ points. This means that
\beq 
\vec x = {b\over Q}\vec l ~~,~~ \vec k = {2\pi \over b} \vec n 
\ ,
\eeq
where $\vec l$ is a triplet of integers between $0$ and $Q-1$ and
$\vec n$ is a triplet of integers between $-Q/2$ and $(Q/2)-1$.
We can move between the finite discrete set of points and the
continuous limit using the usual substitutions
\beq
\int d^3x \to \left ( { b\over Q} \right ) ^3 
\sum_{\vec x} ~~,~~ \int
d^3k \to \left ( { 2\pi\over b} \right ) ^3
\sum_{\vec k}
\ .
\eeq
Our convention for the Fourier transform is
\beqn
\nonumber
f(\vec x) &=& \int d^3k e^{i\vec k\cdot \vec x} f(\vec k) = \left
( { 2\pi\over b} \right ) ^3 \sum _{\vec k} e^{i\vec k\cdot \vec
x}f(\vec k)
\\
f(\vec k) &=& \left ( {1\over 2\pi}\right ) ^3 \int d^3x e^{i\vec
k\cdot \vec x} f(\vec x) = \left ( {1\over 2\pi}\right ) ^3 \left
( { b\over Q} \right ) ^3
\sum _{\vec x} e^{i\vec k\cdot \vec
x}f(\vec x)
\ .
\eeqn
We will expand the waterfall field in modes in momentum space,
\beq
\phi(\vec x
,t) = {1\over (2\pi)^{3/2}} \left ({2\pi \over b}\right )^{3/2}
\sum _{\vec k} [c(\vec k) e^{i\vec k \cdot \vec x} u(\vec k,t )
+d^\dag(\vec k) e^{-i\vec k \cdot \vec x} u^*(\vec k,t )]
\ ,
\label{eqn:phi(x)}
\eeq
which with Eq.~(\ref{eqn:phieq}) gives
\beq
\ddot u(\vec k,N) + 3 \dot u(\vec k,N)+e^{-2N} \tilde k^2 u(\vec
k,N) = \mu_\phi ^2( 1-e^{-\tilde \mu_\psi ^2N})u(\vec k,N) \ ,
\label{eqn:modefn}
\eeq
where $\tilde k = {|\vec k| \over H}$ and an overdot denotes a
derivative with respect to the time variable $N=Ht$.

\subsection{Solution of the mode function}

\subsubsection{Early time behavior}

At asymptotically early times the $\tilde k^2$ term dominates
over the mass term provided that $\tilde \mu_\psi^2 < 2$.  For
$r=2$ this is the case for $\mu_\psi < \sqrt{2}$, and for $r=4$
it holds for $\mu_\psi < \sqrt{5/4}$.  These inequalities will
hold throughout the parameter space of the models that we will
examine, so we can neglect the mass term for $N \to -\infty$. We
then define a new function, following \cite{guthpi2} as
\beq
u (\vec k,N) ={1\over 2 } \sqrt {\pi \over H} e^{-3N/2} Z(z)
~~,~~ z=\tilde k e^{-N}
\ .
\eeq
Neglecting the mass term in Eq.~(\ref{eqn:modefn}), we find
\beq
z^2 {d^2Z\over dz^2} + z {dZ\over dz} + \left (z^2 - {9\over 4}
\right ) Z=0\ ,
\eeq
which is the equation for a Bessel function of order 3/2. At
early times the solution should look like a harmonic oscillator
in its ground state, or equivalently the ground state of a free
field in flat space, which is composed of negative frequency
complex exponentials. This choice of initial conditions is the
well known Bunch--Davies vacuum.  For a review of scalar field
quantization in de Sitter space and the corresponding vacuum
choice see for example Ref.~\cite{mukhanov}. At early times the
solution is given by
\beq 
u \sim {1\over 2} \sqrt {\pi \over H} e^{-3N/2} H^{(1)} _{3/2}(z)
 \ ,
\eeq
where
\beq
H_{3/2}^{(1)}(z) = -\sqrt {2\over \pi z} e^{iz} \left (1-{1\over
iz} \right)
\eeq
is a Hankle function, a linear combination of Bessel functions. 
(The phase is arbitrary, and the normalization is fixed by
insisting that the field and the creation and annihilation
operators obey their standard commutation relations.) Rewriting
the original mode equation in terms of the new variable $z$, it
simplifies to
\beq
{\partial^2 u \over \partial z^2} -{2 \over z} {\partial u \over
\partial z} +u = {\mu_\phi ^2 \over z^2} \left [ 1- \left (
{z\over \tilde k} \right )^{\tilde \mu_\psi^2} \right ] u
\ .
\eeq
The $\vec k =0$ mode is not captured by the procedure described
here and is presented in detail in Appendix \ref{AppZero}.

\subsubsection{General Solution}

We will now examine the general solution in a form that will be
more appropriate for the numerical calculations that we have to
perform. We can write the solution as
\beq
u(\vec k,t) = {1\over \sqrt{2\tilde k H}} R(\vec
k,t)e^{i\theta(\vec k,t)}
\eeq
and the differential equation separates in real and imaginary
parts
\beqn
 \ddot R - R\dot \theta ^2 + 3\dot R + e^{-2N} \tilde k^2 R & = & \mu_\phi ^2( 1-e^{-\tilde \mu_\psi ^2N})R
 \\
2\dot R \dot \theta+ R\ddot \theta +3R\dot \theta & = & 0
\eeqn
Integrating the second equation gives
\beq
\dot \theta = const {e^{-3N} \over R^2}
\eeq
By comparing this with the early time behavior of the analytic
solution
\beq
u\sim {1\over 2H \sqrt {\tilde k}} e^{-N}e^{i\tilde k e^{-N}}
\eeq
the phase equation becomes
\beq
 \dot \theta = - {\tilde k e^{-3N} \over R^2}  
\eeq
while the initial condition for the amplitude is given by the
same asymptotic term to be
\beq
R\to e^{-N}
\eeq
Inserting this expression in the equation for the amplitude
function $R$
\beq
 \boxed { \ddot R - {\tilde k^2 e^{-6N} \over R^3} + 3\dot R + e^{-2N} \tilde k^2 R = \mu_\phi ^2( 1-e^{-\tilde \mu_\psi ^2N})R  }
 \eeq

\subsubsection{A closer look at the mode behavior}

\label{subsec:mode-behavior}

Let us rewrite the equation of motion (Eq. \ref{eqn:modefn}) in a
way that makes the time dependence of the solution more
transparent
\beq 
\ddot u_k(t) + 3 \dot u_k(t) +\mu_{eff}^2 =0 ~~,~~ \mu_{eff}^2(k)  = \tilde k^2 e^{-2N} + \mu_\phi^2 e^{-\tilde \mu_\psi^2 N} -\mu_\phi^2
\label{eqn:meff}
\eeq
We can distinguish different time windows with different behavior
of the mode functions, based on the effective waterfall field
mass. We will list these time windows here and then proceed to
examine them one by one.
\begin{enumerate}
\item $N\ll 0$, many efolds before the waterfall transition, in the asymptotic past
\item $N_{dev}(k)<N<0$, a few efolds before the transition, where $N_{dev}(k)$ is the time at which a mode starts deviating significantly from the $e^{-N}$ behavior, in particular starts decaying faster.
\item $0<N< N_{tr}(k)$ a few efolds after the transition, where $N_{tr}(k)$ is the time at which each mode starts growing. 
\item $N \gg 0$, the asymptotic future
\end{enumerate}
Now let us look at each of those time scales more closely. The
asymptotic past is well described in the previous section and we
see that all modes decay like $e^{-N}$. More precisely their
magnitude behaves as $|u_k| \sim \sqrt {1\over 2k} e^{-N}$. 
The first time scale $N_{dev}$ appears only for low wavenumbers.
For $N<0$ we can keep only two of the three terms in the
effective mass. Since $\mu_\phi^2 e^{-\tilde \mu_\psi^2 N}
>\mu_\phi^2$ we will drop the $\mu_\phi^2$ term, leaving the
effective mass as $\mu_{eff}^2(k) = \tilde k^2 e^{-2N} +
\mu_\phi^2 e^{-\tilde \mu_\psi^2 N} $. The time at which the two
dominant terms become equal is
\beq
N_{dev}(k)={2\over 2-\tilde \mu_\psi ^2} \log \left ( {\tilde k
\over \mu_\phi } \right )
\eeq
For $\tilde k \ge \mu_\phi$ this time is not negative, hence we
cannot drop $\mu_\phi^2$ and our analysis fails. This transition,
which happens only for $\tilde k < \mu_\phi$ signals a deviation
of the behavior of the modes, which do not evolve as $e^{-N}$,
but instead decay faster. 

Next we move to the actual waterfall transition time for each
mode, which happens when the effective squared mass changes sign
and becomes negative, or
\beq 
\tilde k^2 e^{-2N} = \mu_\phi^2\left ( 1- e^{-\tilde \mu_\psi^2 N} \right )
\label{eqn:tran}
\eeq
We will approximate the right hand side of the above equation
with a piecewise linear function as follows
\beq \mu_\phi^2\left ( 1- e^{-\tilde \mu_\psi^2 N} \right ) =\left\{
     \begin{array}{lr}
       \mu_\phi^2 \tilde \mu_\psi ^2 N & : N < 1/ \mu_\psi^2\\
       \mu_\phi^2 & : N > 1/ \mu_\psi^2
     \end{array}
   \right.
   \eeq
For $\tilde k < \mu_\phi e^{1/\mu_\psi^2}$ the solution is found
on the first branch and is
\beq
N_{tr}(k)={1\over 2} W \left ( {2\tilde k ^2 \over \mu_\phi
\tilde \mu_\psi ^2} \right )
\eeq
where $W$ is known as the Product Logarithm, or Lambert W
function and is defined as the solution to the equation
$z=W(z)e^{W(z)}$. For small values of the wavenumber we can write
the solutions as a Taylor series in $\tilde k$
\beq
N_{tr}(\tilde k \ll \sqrt {\mu_\phi \tilde \mu_\psi}) = {\tilde
k^2 \over \mu_\phi^2 \tilde \mu_\psi^2} + O(\tilde k^4)
\eeq
For $\tilde k > \mu_\phi e^{1/\mu_\psi^2}$ we operate on the
second branch and the transition time for each mode is
\beq
N_{tr}(k) = \log \left ( {\tilde k \over \mu_\phi } \right )
\eeq

The behavior of the modes after the transition is different for
different ranges of the timer field mass.  If we consider the
late time behavior of the mode equation, we can see two
timescales introduced by the time dependent exponential
coefficients. One is $O(1)$ and the other $O(1/\tilde \mu_\psi
^2)$. We distinguish two cases: They can both be O(1) or the
second one can be larger than the first. The first timescale
defines the time at which the equation becomes k-independent,
meaning that all modes behave (grow) in the same way. The second
time scale defines the time, after which the equation becomes
time independent, meaning that after that all modes behave as
pure exponentials. 

Let us first deal with the case of $\tilde \mu_\psi \ll 1$
meaning that the second time scale is much larger than the first
one. Between the two timescales, that is $1< N < -1/\tilde
\mu_\psi ^2$, the equation is independent of $\tilde k$
\beq
 \ddot R + 3\dot R  = \mu_\phi ^2( 1-e^{-\tilde \mu_\psi ^2N})R  
 \eeq
Since the evolution of the exponential term on the right hand
side is by far slowest than the other timescales in the problem,
we will treat $1-e^{-\tilde \mu_\psi ^2N}$ adiabatically. This
leads immediately to the solution
\beq
R = R_0 e^ {\lambda(N) N} ~~,~~ \lambda (N) =
{-3+\sqrt{9+4\mu_\phi ^2( 1-e^{-\tilde \mu_\psi ^2N})} \over 2}
\label{lambda(N)}
\eeq
After a long time, this would mathematically settle to
\beq
\lambda_0= {-3+\sqrt{9+4\mu_\phi ^2} \over 2} \ .
\label{eqn:lambda0}
\eeq
However, this is far beyond the time when inflation will have
ended, hence it would physically never have time to happen (plus
it is well outside the validity of our constructed potential).

\begin{figure}
\centering
\begin{tabular}{cc}
\includegraphics[width=3in]{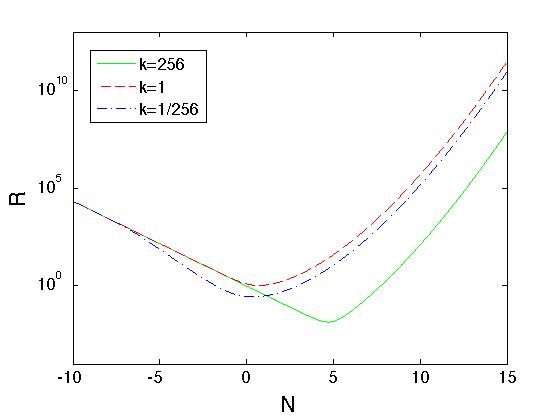} &
\includegraphics[width=3in]{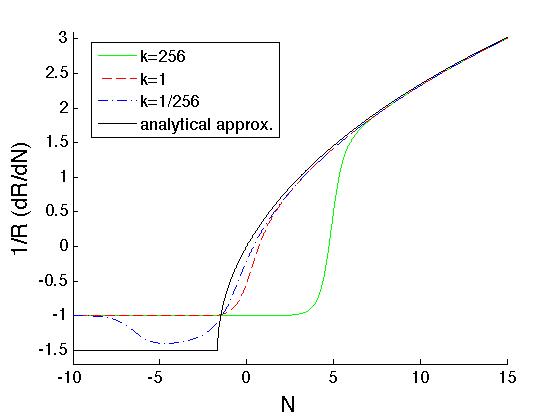}
\end{tabular}
\caption{Mode functions for different comoving wavenumbers as a
function of time in efolds. The model parameters are
$\mu_\psi={1\over 10}$ and $\mu_\phi=10$. We can see the modes
following our analytic approximation for the growth rate. Our
analysis gives $N_{dev}(1/256) \approx -7.9$ and
$N_{tr}(256)\approx 4.76$, which are very close to the values
that can be read off the graph.}  \label{fig:modes_1}
\end{figure}

Let us choose $\mu_\phi = 10$ and $\tilde \mu_\psi = 1/10$ to
demonstrate our analysis.  Some characteristic mode functions are
presented in Fig. \ref {fig:modes_1}

We see that all modes behave similarly at late times, independent
of their wave-number, as they should based on our late time
analysis. Specifically, we can plot the ratio of the time
derivative of each mode to its magnitude, as in Fig.
\ref{fig:modes_1}. We call this the growth rate $\lambda \equiv
\left ( {\dot R_k \over R_k}\right )$. We can see both phenomena.
First, after $N \approx 6$ the modes behave identically. Second,
the behavior of the mode approaches that of an exponential
function (whose logarithm is a constant), but at a slower rate.
In this example time needs to go on for several hundreds of
efolds for the growth rate to set to a constant, which is
calculated to be $\lambda (t \to \infty) = 8.6119$ for $\mu_\phi
= 10$ and $\mu_\psi = 1/10$.

It is important to test our analytical approach to the late time
behavior of the growth rate of mode function. As seen in Fig.
\ref{fig:modes_1} once the mode functions evolve in a $k$ -
independent way, our simple analytical estimate for their growth
rate is accurate to within a few percent, which gives us a very
accurate expression for the growth rate and leads to the terms
evolving as $u\sim e^{\lambda(t) t}$, where the time-dependent
growth rate $\lambda(t)$ is slowly changing.

As a test of our analysis, we can calculate the two important
transition times $N_{dev}(1/256) =-7.9005$ and
$N_{tr}(256)=4.76457$. We see that the calculated values agree
very well with the behavior of the plotted modes.

\begin{figure}
\centering
\begin{tabular}{cc}
\includegraphics[width=3in]{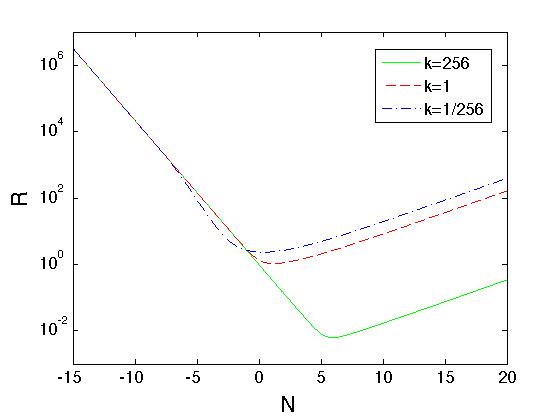} &
\includegraphics[width=3in]{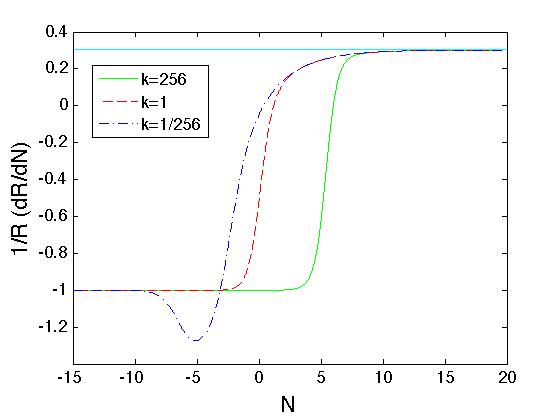}
\end{tabular}
\caption{Mode functions for different comoving wavenumbers as a
function of time in efolds. The model parameters are
$\mu_\psi={1\over 2}$ and $\mu_\phi=1$. The horizontal line
corresponds to the asymptotic value of the growth factor
$\lambda$. We can see how the mode functions reach their
asymptotic behavior after $10$ efolds. Our analysis gives
$N_{dev}(1/256) \approx -7.84$ and $N_{tr}(256)\approx 5.4$,
which are very close to the values that can be read off the
graph.}
 \label{fig:modes_2}
\end{figure}

Let us briefly examine the situation where $\tilde \mu_\psi ^2
\le 2$. In this case the mode equation becomes k-independent and
time independent at about the same time, that is a few e-folds
after the waterfall transition. We choose $\mu_\psi = {1\over 2}$
and $\mu_\phi=1$ and plot the results in Fig. \ref{fig:modes_2}.
It is clear that the modes become both $k$-independent and pure
exponential (having a constant growth rate) at about the same
time ($N\approx 10$). The asymptotic growth rate in this case is
$\lambda (t \to \infty) = 0.3028$.

Again we can calculate the two important transition times
$N_{dev}(1/256) =-7.8377$ and $N_{tr}(256)=5.39554$, which agree
once more with the behavior of the plotted modes.

%%%%%%%%%%%%%%%%%%%%%%
%%%%%%%%%%%%%%%%%%%%%%
%%%%%%%%%%%%%%%%%%%%%%

\section {Supernatural Inflation models}

It is interesting to make contact between our abstract model and
specific potentials inspired form particle theory. In general
inflation models require small parameters in order to ensure slow
roll inflation and produce the correct magnitude of density
perturbations. It was shown in \cite{sol1,sol2} that
supersymmetric theories with weak scale supersymmetry breaking
can give models where such small parameters emerge "naturally" as
ratios of masses already in the theory. We will not go into the
details of such theories, but instead give the forms of the
constructed potentials and use them as an application of our
formalism.
\beq
V=M^4 \cos^2(\phi/\sqrt 2 f) + {m_\psi ^2 \over 2} \psi ^2 +
{\psi ^4\phi^2 + \phi^4 \psi ^2 \over 8 {M'} ^2}
\label{SUSYModel1}
\eeq
for what we will call model 1 and will be the primary focus of
this work and
\beq
V=M^4 \cos^2(\phi/\sqrt 2 f) +{m_\psi ^2 \over 2} \psi ^2 +
\lambda^2 {\psi ^2 \phi ^2 \over 4}
\label{SUSYModel2}
\eeq
which we will call model 2. 

The first model can be taken with $M'$ at one of three regions:
the Planck scale, the GUT scale or an intermediate scale ($\sim
10^{10} ~ GeV$). At each scale the rest of the parameters are
adjusted accordingly to produce sufficient inflation and agree
with CMB data. 

We will approximate the potential with a pure quadratic one with
a time varying waterfall mass, of the form
\beq
V(\phi, \psi) = V_0 -m_0 ^2 \left [ 1- \left ({\psi \over \psi_c}
\right ) ^r \right ] |\phi|^2 + m_\psi ^2 \psi^2
\eeq
where $r=4$ for model 1 and $r=2$ for model 2, as can be easily
seen by the form of the interaction terms in both cases. The
correspondence between the exact SUSY potential and our quadratic
counterpart is shown in Table
\ref {tab:susy}.

\begin{figure}
\centering
\begin{tabular}{cc}
\includegraphics[width=3in]{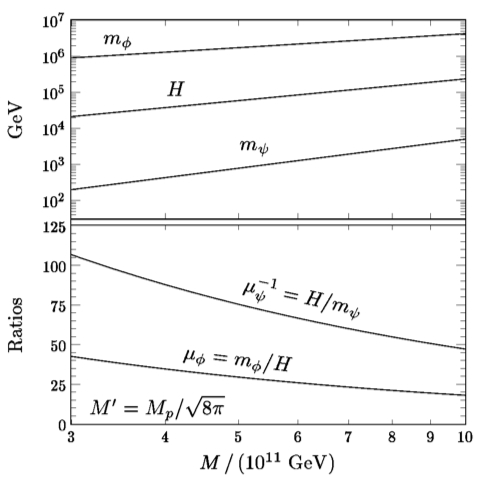}&
\includegraphics[width=3in]{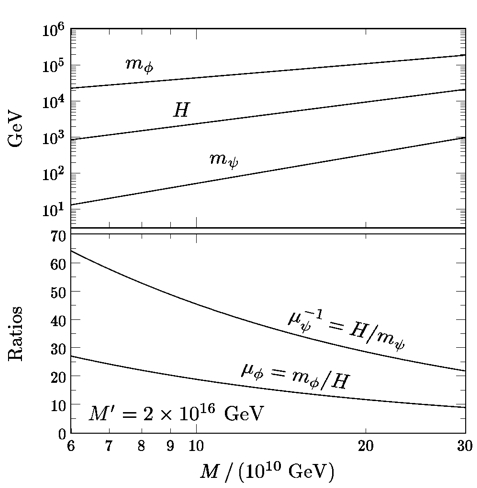} 
\\
\includegraphics[width=3in]{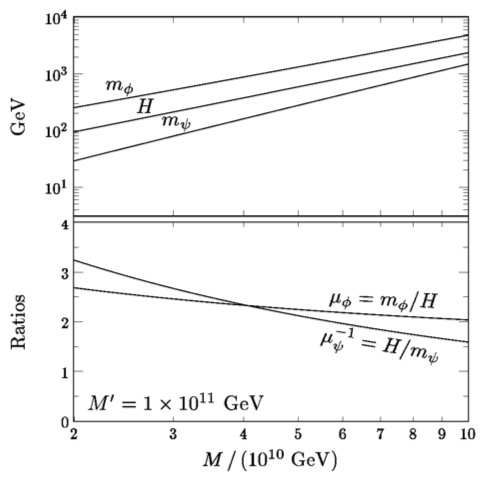}&
\includegraphics[width=3in]{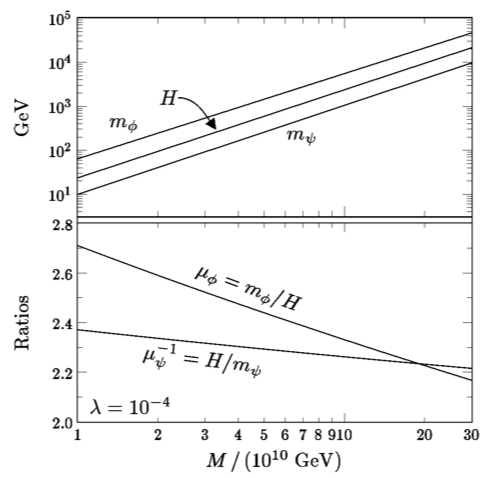} 
\end{tabular}
   \caption{Parameter space for the two supernatural inflation
models. The bottom right corner shows the parameter for model 2,
while the other three show parameters for model 1, for different
ranges of the mass scale $M'$ }
   \label{fig: model}
\end{figure}

The parameters of the two models are restricted to fit CMB data,
as shown in Fig. \ref{fig: model}.

\begin{table}
    \begin{tabular}{|c|c|c|c|c|c|}
        \hline
     Quadratic Approximation & ~$V_0$~ &~ $m_0$ ~ & ~$r$~&~
$\psi_c$ ~ & ~$m_\psi$~ \\
        \hline
     SUSY Model 1 & ~$M^4$ & ~${M^2 \over \sqrt 2 f}$ ~ & ~$4$~&
~${M \sqrt {2M'}\over \sqrt f} $~& ~$m_\psi$~ \\
        \hline
     SUSY Model 2 & ~$M^4$~ & ~${M^2 \over \sqrt 2 f}$ ~ & ~$2$
~& ~${\sqrt 2 M^2 \over f \lambda}$ ~& ~$m_\psi$~\\
         \hline
    \end{tabular}
    \caption{ Parameters of SUSY models and their counterparts in our quadratic approximation}
    \label{tab:susy}
\end{table}

To lowest order, in this potential dominated model, the Hubble
parameter is constant and equal to
\beq
H=\sqrt{8\pi \over 3} {M^2 \over M_p} = \sqrt{8\pi \over 3}
{\sqrt {V_0} \over M_p}
\eeq

\subsection{End of Inflation}

In our simplified quadratic model inflation will never end. The
waterfall field will roll forever down its tachyonic potential. 
However, we shall not forget that this is a mere Taylor expansion
of more realistic potentials, which have a well defined minimum.
We will use the supersymmetric potentials of \cite{sol1,sol2} as
a concrete example to connect our purely quadratic potential to
ones with more realistic shapes. In these supersymmetric models
the potential has a cosine-like form and the minimum occurs at
${\phi \over \sqrt 2 f}= {\pi \over 2}$, where the inflaton will
oscillate, terminating inflation and giving rise to (p)reheating.
By making contact between the parameters of our potential and the
physical parameters of the actual supersymmetric models, we can
estimate the field value at which inflation ends.

There are two strategies for defining $\phi_{\rm end}$, the field
value at which inflation ends. We can either pretend that the
quadratic potential can be followed up to the end field value of
the corresponding SUSY potential, or we can choose to end our
calculation when the quadratic potential departs significantly
from the actual SUSY potential that we are trying to approximate.

In the first case the end field value is at $\phi_{\rm end} =
{f\pi / \sqrt 2}$.  To calculate the end field value for the
latter case we will note that the cosine potential is accurately
approximated by a quadratic as long as $\phi / f \ll 1$. We will
call this ratio $\epsilon$ and in this case we will end our
calculations when $\epsilon$ ceases being small. We can write
these two cases in a unified manner, as
\beq
 \phi_{\rm end}
= \epsilon f
\eeq
where $\epsilon ={\pi / \sqrt 2}$ if we follow the quadratic
potential all the way to the field value corresponding to the
minimum of the SUSY potential and $\epsilon < 1$ if we stop our
calculation at the point where the quadratic potential deviates
significantly from the supersymmetric one.

Using the values of the parameters taken from the supersymmetric
models, we can estimate the end field value to be
\beq
\phi_{\rm end} \sim  \epsilon~ 10^{15} H
\eeq
within one or two orders of magnitude for all cases of models
considered in \cite{sol1,sol2}.

We will be using field values of this order of magnitude in our
numerical calculations, whether we are dealing with the
supersymmetric potentials or not. We will however examine the
effects of changing the end value of the field and show that it
is minimal, easily understandable, and calculable.

%%%%%%%%%%%%%%%%%%%%%%%%%%%%%%%%%%%%
%%%%%%%%%%%%%%%%%%%%%%%%%%%%%%%%%%%%
%%%%%%%%%%%%%%%%%%%%%%%%%%%%%%%%%%%%
%%%%%%%%%%%%%%%%%%%%%%%%%%%%%%%%%%%%

\section{ Perturbation theory basics}

\subsection{Time delay formalism}

The time delay formalism provides an intuitive and
straightforward way to calculate primordial perturbations. Its
basic principle is that inflation ends at different places in
time at different times, due to quantum fluctuations. This leads
some of the regions of the universe to have inflated more than
others, creating a difference in their densities. The time-delay
formalism was first introduced by Hawking \cite{hawking} and by
Guth and Pi \cite{guthpi1}, and has recently been reviewed in
Ref.~\cite{guthsolvay}.

We will briefly describe the method here for the case of a single
real scalar field.  The universe is assumed to be described by a
de-Sitter space-time, since the Hubble parameter is taken to be a
constant.  The equation of motion for the scalar field $\phi(\vec
x,t)$ is
\beq
\ddot \phi + 3H \dot \phi = - {\partial V \over \partial \phi} + {1\over a(t) ^2 } \nabla^2 \phi
\eeq
where the last term is suppressed by an exponentially growing
quantity, so at late times it becomes negligible. We will omit
the last term from now on.

We call the homogenous (classical) solution $\phi_0(t)$ and write
the full solution, including a space dependent small perturbation
$\delta \phi \ll \phi_0$ as
\beq
\phi(\vec x, t ) =\phi_0( t ) +\delta \phi(\vec x, t ) 
\eeq
Plugging this into the equation of motion and working to linear
order in $\delta \phi$ one can show that the quantity $\delta
\phi$ obeys the same differential equation as $\dot \phi_0$.
Furthermore the presence of a damping term implies that any two
solutions approach a time independent ratio at large times. Thus,
at large times we have (to first order in $\delta \tau $)
\beq
\delta \phi (\vec x,t) \to - \delta \tau (\vec x) \dot \phi_0(t) \Rightarrow \phi (\vec x,t) \to \phi_0(t - \delta \tau (\vec x))
\eeq
This is the formulation of the intuitive picture of the time
delay method.

\subsection{Randall-Soljacic-Guth approximation}

The usual calculation of density perturbations in inflation
considers small quantum fluctuations around a classical field
trajectory. In the case of hybrid inflation such a classical
trajectory does not exist, since classically the field would stay
forever on the top of the inverted potential. It is the quantum
fluctuations that push the field away from this point of unstable
equilibrium. One way to overcome this difficulty is to consider
the RMS value of the field as the classical trajectory. This was
done for example in \cite{sol1} and \cite{sol2} and is a
recurring approximation in the study of hybrid inflation. 

Using the Bunch-Davies vacuum in the definition of the RMS value
of the waterfall field $\phi_{\rm rms} = \sqrt { \left < 0 |
\phi(x,t) \phi^*(x,t) | 0 \right >}$ it it straightforward to
calculate it using the mode expansion
\beq
\phi_{\rm rms}^2(t) = {1\over b^3} \sum_{\vec k ~, ~\vec k'} e^{i(\vec k -\vec k')\cdot x}
\left < 0|    (c_{\vec k} u _{\vec k} +       d^\dagger_{-\vec k} u ^*_{-\vec k } ) (c^\dagger_{\vec k'} u^* _{\vec k'} +    d_{-\vec k'} u_{-\vec k' } ) |0 \right > =  {1\over b^3}  \sum_{\vec k} |u_k(t)|^2 
\eeq
The mean fluctuations are measured by
\beq \Delta \phi (\vec k) = \left [ 
\left( {k \over 2 \pi} \right)^3
\int d^3x e^{i\vec k \cdot \vec x} \left < \phi(x)
\phi^*(0) \right > \right ] ^{1/2} =\left [ \left ({k\over 2\pi
}\right)^3 |u_{\vec k} |^2\right ]^{1/2}
\eeq
resulting in what will be called the RSG approximation for the
time delay field
\beq
 \Delta \tau_{RSG} (\vec k) \approx {\Delta \phi (\vec k,t) \over \dot \phi_{\rm rms}} 
= \left ({kb\over 2\pi }\right)^{3/2} |u_{\vec k} |{ \sqrt
{\sum_{\vec k} {|u_{\vec k} (t)|^2 }} \over \sum _{\vec k} {{
\dot u_{\vec k} (t) u_{\vec k} (t) }} }
\label{RSG}
\eeq
There is an important comment to be made about the quantum
mechanical nature of these density perturbations. In regular
models of inflation quantum perturbations are scaled by $\hbar$.
We can think of them as modes with initial conditions that are of
the order of $\hbar$. The classical trajectory on the other hand
does not have any quantum mechanical origin, hence does not scale
with $\hbar$. This means that in the limit of $\hbar \to 0$ the
perturbations vanish, as one would expect will happen if one
could "switch off" quantum mechanical effects. 

In the case of hybrid inflation on the other hand, what we call
the classical trajectory (be it the RMS value or something else)
is comprised of modes that originated as quantum fluctuations,
hence is scaled by $\hbar$ itself. This means that even in the
limit of $\hbar \to 0$, the density perturbations in hybrid
inflation remain finite! By explicitly restoring $\hbar$ in the
formulas of the paper, the reader can formally arrive to the same
conclusion.

\begin{figure}
\includegraphics[width=3in]{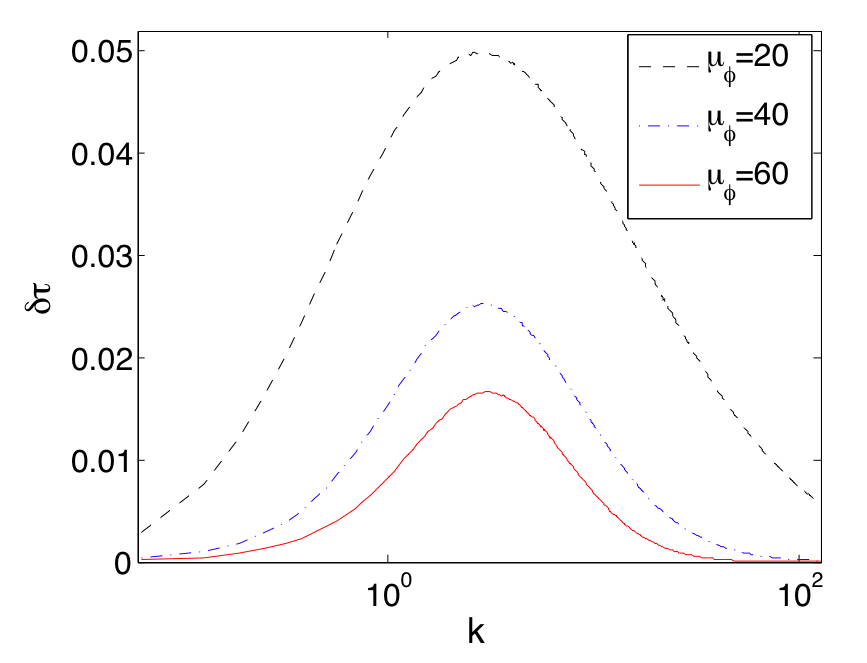}
\caption{The time delay field calculated using the RSG formalism. The end time was taken to be $15$ e-folds after the waterfall transition and $\mu_\psi ={1\over 20}$ for all curves, while we varied $\mu_\phi$. }
 \label{fig:rsg}
\end{figure}

Some plots of the time delay field calculated using the RSG
approximation are shown in Fig. \ref {fig:rsg}. The reduced mass
of the timer field was taken to be $\mu_\psi ={1\over 20}$ while
we varied the waterfall field mass. We fixed the time at which
inflation ended to be $15$ e-folds after the waterfall
transition. 

%%%%%%%%%%%%%%%%%%%%%%%%%%%%%%
%%%%%%%%%%%%%%%%%%%%%%%%%%%%%%

\section{%
Calculation of the Time Delay Power Spectrum }

The usual method to calculate the primordial perturbation
spectrum would involve either making some approximations (more or
less similar to the RSG) or using a Monte Carlo simulation. The
former suffers from the lack of a classical trajectory that
invalidates the usual perturbation method, while the latter would
be computationally costly in three spatial dimensions. We will
therefore implement an alternate method that reduces the
calculation of the spectrum of the time delay field to the
evaluation of a two dimensional integral and does not need a
classical trajectory to do so.

As discussed at the end of Sec.~\ref{subsec:mode-behavior}, the
behavior of the mode functions at asymptotically late times ($t
\to \infty$) is given by
\beq
u(\vec k ,t\to \infty) \sim e^{
\lambda_0  
t} u(\vec k)
\ ,
\eeq
where $\lambda_0$ is given by Eq.~(\ref{eqn:lambda0}). If we
define for all times
\beq
\lambda(t) \equiv {\dot \phi_{\rm rms}(t) \over \phi_{\rm rms}(t)} =
{\sum _{\vec k} {R(\vec k,t) \dot R(\vec k,t) \over 2 |k|} \over
\sum_{\vec k} {|R(\vec k ,t)|^2 \over 2|k|}}
\ , 
\eeq
then at late times $\lambda(t) \to \lambda_0$. Since $\lambda(t)$
changes very slowly, we can take it as a constant around the time
of interest.

To discuss fluctuations in the time at which inflation ends, we
begin by defining $t_0$ as the time when the rms field reaches
the value $\phi_{\rm end}$, which we have chosen to define the
nominal end of inflation:
\beq
\phi_{\rm rms} ^2(t_0) = \phi^2_{\rm end} 
\ .
\eeq
Since at late times all modes, to a good approximation, grow at
the same exponential rate $\lambda(t)$, we can express the field
$\phi(\vec x,t)$ at time $t = t_0 + \delta t$ in terms of the
field $\phi(\vec x, t_0)$ by
\beq
|\phi(\vec x,t)|^2 = |\phi(\vec x,t_0)|^2 e^{2\lambda \delta t}
\ .
\eeq
If $t$ is chosen to be the time $t_{\rm end}(\vec x)$ at which
inflation ends at each point in space, then $\phi\bigl(\vec
x,t_{\rm end}(\vec x)\bigr) = \phi_{\rm end} =
\phi_{\rm rms}(t_0)$, and the above equation becomes
\beq
\phi_{\rm rms}^2(t_0) = |\phi(\vec x,t_0)|^2 e^{2\lambda  \delta
t} \ ,
\eeq
which can be solved for the time delay field $\delta t(\vec x) =
t_{\rm end}(\vec x) - t_0$:
\beq
\delta t(\vec x) = {-1\over 2\lambda} \log \left ({ |\phi(\vec
x,t_0)|^2 \over \phi_{\rm rms}^2(t_0) }\right )
\ .
\eeq
Rescaling by the rms field
\beq
\tilde \phi (\vec x ,t) 
\equiv
{\phi (\vec x ,t)\over \phi_{\rm rms}(t)}
\ ,
\eeq
we can write
\beq
\delta t(\vec x) = {-1\over 2\lambda} \log { |\tilde \phi (\vec x,t_0)|^2 }
\ .
\eeq
Using this expression, we can write the two-point function of the
time delay field as
\beq
\left < \delta t (\vec x)\delta t 
(\vec 0)
\right > = {1\over 4\lambda^2} \left < \log|\tilde \phi(\vec
x,t_0)|^2 \log |\tilde \phi(
\vec 0
,t_0)|^2\right >
\ ,
\eeq
which can be evaluated, since the probability distributions are
known.  To continue, we can decompose the complex scalar field in
terms of the real fields $X_i$:
\beq
\tilde \phi(\vec x,t) = X_1 +i X_2 ~~ , ~~ \tilde \phi(
 \vec 0
,t) = X_3 +i X_4
\ .
\eeq
The average value of a function $F$ of a random variable $X$ with
probability distribution function $p(X)$ is given by
\beq
\left < F[X] \right > = \int dX p(X) F[X]
\ .
\eeq
Since this is a free field theory, we can take the four random
variables $X_i (\vec x)$ to follow a joint Gaussian distribution
with
\beq
p(X) = {1\over (2\pi)^2 \sqrt {\det(\Sigma)}} \exp \left (
-{1\over 2} X^T \Sigma^{-1} X \right ) ~~,~~ \Sigma_{ij} =
\left < X_i X_j \right >\ .
\eeq
A function of the $X_i's$ then has the expected value
\beq
\left < F[X] \right > = \int 
\prod_{i=1}^4 
dX_i {1\over (2\pi)^2 \sqrt {\det(\Sigma)}} \exp \left ( -{1\over
2} X^T
\Sigma^{-1}
X \right ) F[X]
\ .
\eeq
The new fields $X_i$ can be written in terms of the original
complex field $\phi$ as
\begin{eqnarray}
\nonumber
X_1 &=& {1\over 2} \left [ \tilde \phi(\vec x) + \tilde \phi^*(\vec x) \right ]~,~X_2 = {1\over 2i} \left [ \tilde \phi(\vec x) - \tilde \phi^*(\vec x) \right ]
\\
X_3 &=& {1\over 2} \left [ \tilde \phi(
 \vec 0
) + \tilde \phi^*(
\vec 0
) \right ]~,~X_4 = {1\over 2i} \left [ \tilde \phi(
\vec 0
) - \tilde \phi^*(
\vec 0
) \right ]
\end{eqnarray}
The components of the variance matrix $\Sigma$ can be easily
calculated using the commutation relations for the creation and
annihilation operators in $\phi(\vec x, t)$, from
Eq.~(\ref{eqn:phi(x)}). Due to the high degree of symmetry the
matrix itself has a very simple structure:
\beq
\Sigma =  \left (
\begin{array}{cccc}
{1\over 2} & 0 & \Delta & 0 \\ 0 & {1\over 2} & 0 & \Delta\\
 \Delta & 0& {1\over 2} & 0\\
0 & \Delta & 0 & {1\over 2}
 \end{array}
 \right )
\ ,
\eeq
where
\beq
\Delta (\vec x,t_0) = \left < X_1 X_3 \right > = \left < X_2 X_4
\right > = {1 \over 2} \left < \phi^*(\vec x,t_0) \phi(
\vec 0
,t_0) \right > = {1\over 2b^3} \sum _{\vec k} |\tilde u (\vec
k,t_0)|^2 e^{i\vec k \cdot \vec x} \ ,
\eeq
where
\beq
\tilde u (\vec k,t) = {u (\vec
k,t)\over \phi_{\rm rms}(t)} \ .
\eeq
Since $\tilde u(\vec k ,t)$ actually depends only on the
magnitude of the wavenumber, because of the isotropy of the
problem, we can do the angular calculations explicitly in
$\Delta$ and leave only the radial integral to be calculated
numerically. Then
\beqn
\nonumber
\left < \delta t (\vec x) \delta t 
(\vec 0)
\right > &=&{1\over 4 \lambda ^2} 
{1 \over (2\pi)^2 [{1\over 4} - \Delta^2]} \int dX_1dX_2dX_3dX_4
\log(X_1^2+X_2^2) \log(X_3^2+X_4^2) 
\\
&&  %
\hskip -20pt \times \exp 
\left \{ -{ 1\over 4 [{1\over 4} - \Delta^2] }
\Bigl[X_1^2+X_2^2+X_3^2+X_4^2-4(X_1X_3+X_2X_4)\Delta\Bigr] \right \}
\ .
\label{basic2pt}
\eeqn
Changing to polar coordinates
\begin{eqnarray}
\nonumber
&&X_1=r_1 \cos\theta_1~,~X_2=r_1 \sin\theta_1
\\
&&X_3=r_2 \cos\theta_2~,~X_4=r_2 \sin\theta_2
\ ,
\end{eqnarray}
the integral becomes
\beqn
\nonumber
\left < \delta t(\vec x) \delta t (\vec 0) \right > &=& {2 \over \pi
\lambda^2 (1 - 4 \Delta^2)} \int _0 ^{2\pi} d\theta \int _0
^\infty r_1dr_1 \int_0^\infty r_2 dr_2 \log(r_1) \log (r_2)
\\
&& \times \exp \left[ - {r_1^2+r_2^2-4\Delta r_1 r_2 \cos \theta
\over 1 - 4\Delta^2 }\right]
 \ ,
\eeqn
where we redefined the angular variables as $\theta =
\theta_1-\theta_2$ and $\tilde \theta = \theta_1+\theta_2$ and
integrated over $\tilde \theta$.
Changing also the radial variables
\beq
r_1=r \cos \phi~,~r_2=r \sin\phi
\ ,
\eeq
\begin{eqnarray} 
\nonumber
\left < \delta t(\vec x) \delta t(\vec 0) \right > &=& {1 \over \pi
\lambda^2 (1-4 \Delta^2)} \int _0 ^{2\pi} d\theta \int _0
^{\pi\over 2} d\phi \,  \sin 2 \phi \int_0 ^\infty dr\,
r^3 \, \log(r\,\cos\phi) \log(r\,\sin\phi)
\\
&& \times \exp \left[ - { (1-2 \Delta \sin 2 \phi \,
\cos\theta) \, r^2 \over 1 - 4 \Delta^2} \right]
\ .
\label{eqn:delta-tau-2pt}
\end{eqnarray}
The radial integration can be performed analytically
\begin{eqnarray}  
&&\int_0 ^\infty dr~r^3~  \log(ar) 
\log(br) 
e^ { -c r^2}= 
\\
\nonumber
 &&{1\over 8c^2}  \left [(\gamma-2)\gamma +{\pi^2\over 6}- 2\log(ab) (\gamma -1 + \log(c) ) + 4 \log(a) \log(b) + \log(c) (2\gamma -2+\log( c) )\right ]
\end{eqnarray} 
where $a=\cos\phi$, $b=\sin\phi$, $c= { 1\over (1 - 4 \Delta^2)}
(1-2 \Delta
\, \sin 2 \phi \, \cos\theta)$
and $\gamma$ is the Euler constant $\gamma \approx 0.57721$.

Finally, the spectrum of the time delay field is defined by
\beq
\delta \tau (\vec k)=\left[ \left( {k \over 2 \pi} \right)^3 \int
d^3 x \, e^{i \vec k \cdot \vec x} \left < \delta t(\vec x)
\delta t(\vec 0) \right> \right]^{1/2} \ .
\label{eqn:powerspec}
\eeq

Calculation in the two limiting cases $x \to 0 $ and $x \to
\infty$ (or $x\to b$ in our case) can be done analytically.

\begin{enumerate}
 \item For $x \to 0 $ several terms in the integral diverge, since $\Delta \to {1\over 2}$. In this case we have only two degrees of freedom instead of four, since we consider a complex scalar field at one point in space. The integral becomes
\begin{eqnarray}
\nonumber
  \left < \delta t(\vec 0) \delta t(\vec 0) \right > &=&  
{1\over 4\lambda ^2} \int {dX_1 dX_2 \over \pi} e^{
-(X_1^2+X_2^2)
}
\log^2(X_1^2+X_2^2) =  
\\
&& = {1\over 4\pi \lambda^2 } \int_0 ^{2\pi} d\theta
\int_0^\infty dr~ r e^{-r^2} \log^2 (r^2) ={1\over 4\lambda ^2} \left ( \gamma ^2 + {\pi^2 \over 6 } \right )  
\ .
 \end{eqnarray}

\item The $x \to \infty$ limit is much easier to handle. We
recognize that $\Delta(\vec x)$ is simply the Fourier transform
of $|u_k|^2$. Since $u_k$ is smooth, $\Delta(x\to \infty) \to 0
$, and therefore $\delta t(\infty)$ is uncorrelated with $\delta
t(\vec 0)$.  Eq.~(\ref{basic2pt}) can be seen to factorize,
giving $\left <
\delta t(\infty) \delta t(\vec 0) \right > = \left < \delta
t(\vec 0)\right >^2$, where
\beqn
\left < \delta t(\vec 0)\right > &=& {-1 \over 2 \pi \lambda} \int d X_1 d X_2 \log(X_1^2+X_2^2) \exp\left[ -(X_1^2 +
X_2^2) \right] \nonumber \\ &=& - {1 \over \pi \lambda} \int_0^{2
\pi} d \theta \int_0^\infty r \, d r \, \log r e^{- r^2}
\nonumber \\ &=& {\gamma \over 2 \lambda \ .}
\eeqn

\end{enumerate}

Combining these results, we see that the probability distribution
for $\delta t(\vec 0)$ has a standard deviation $\sigma =
\sqrt{\left < \delta t(\vec 0)^2 \right> -
\left < \delta t(\vec 0) \right>^2 } = \pi/(2 \sqrt{6} \lambda)$.
While the first limit above is needed for programming the
numerical calculations, since the integral of
Eq.~(\ref{basic2pt}) cannot be numerically evaluated at $\vec
x=\vec 0$, the second limit can be used as a numerical check.

The same method can be applied to the exact calculation of any
higher order correlation functions. Especially the non-Gaussian
part of the power spectrum $f_{NL}$ can be read off from the
momentum space Fourier transform of the three-point correlation
function in position space $\left < \delta t(\vec x_1) \delta
t(\vec x_2) \delta t (x_3) \right > =
\left < \delta t(\vec x_1) \delta t(\vec x_2) \delta t (0) \right
> $. Taking the Fourier transform we can compute $\left < \delta
t(\vec k_1) \delta t(\vec k_2) \delta t (k_3) \right > $, from
which we can extract the properties of the bispectrum.

The form of the three point function in position space is
 \beq
 \left < \delta t(\vec x_1) \delta t(\vec x_2) \delta t (0) \right >  = {-(2\pi)^2 \over \lambda ^3 } \int_0 ^{2\pi} d\gamma_1 d\gamma_2 \int_0 ^\pi \sin\theta d\theta \int_0 ^{2\pi} d\phi F(\gamma_1,\gamma_2,\theta,\phi)
 \eeq
where $F(\gamma_1,\gamma_2,\theta,\phi)$ is a function of four
angular variables. Calculations regarding the form of the
bispectrum will be published elsewhere.

%%%%%%%%%%%%%%%%%%%%%%%%%%%%%%
%%%%%%%%%%%%%%%%%%%%%%%%%%%%%%
%%%%%%%%%%%%%%%%%%%%%%%%%%%%%%

\section{Numerical Results and Discussion}

\begin{figure}
\includegraphics[width=4in]{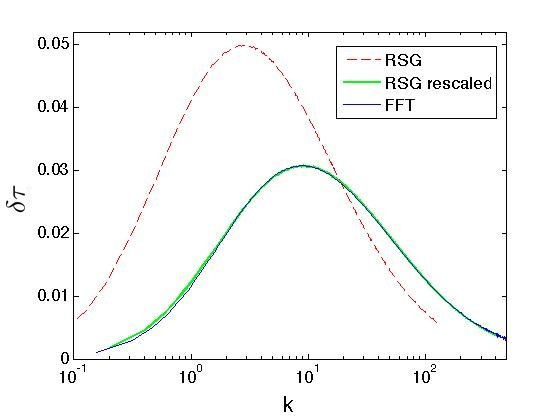}
\caption{Comparison between the RSG and 
FFT methods. The end time was taken to be $15$ e-folds after the
waterfall transition and $\mu_\phi =20$ and $\mu_\psi = 1/20$. 
We can see that the spectrum of the time delay field calculated
in the free field theory agrees very well with the rescaled
version of the RSG approximation $A \, \delta \tau_{RSG} \,
(Bk)$. }
 \label{fig:rsg_di}
\end{figure}

Let us begin by plotting one example of the free field theory
(FFT) calculation of the time delay power spectrum, from
Eqs.~(\ref{eqn:delta-tau-2pt}) and (\ref{eqn:powerspec}), along
with the corresponding curve derived using the RSG approximation,
Eq.~(\ref{RSG}).  We use the sample parameters $\mu_\psi={1\over
20} $ and $\mu_\phi = 20$. Both calculations give a spike, but a
spike of different width, different height and different
position. Let us rescale the RSG result as follows
\beq
\delta \tau _{\rm RSG,rescaled} (k)= A \, \delta \tau_{\rm RSG}
\, (B k) \ ,
\eeq
where $A$ and $B$ are $O(1)$ constants calculated by requiring
the peaks of the FFT and RSG distributions to match in position
and amplitude. The results are plotted in Fig.~\ref {fig:rsg_di}.
We can see that the FFT and RSG curves do not seem similar.
However the rescaled RSG curve seems to follow the FFT curve very
well, as was first noticed by Burgess \cite{kristin}.
Based on our simulations the curves generally tend to agree
better for low wavenumbers, up to and including the peak, and
start deviating after the peak. The rescaling parameters vary
with the field masses chosen and for the particular choice of
Fig.~\ref{fig:rsg_di} were calculated to be $A=0.6152$ and
$B=3.25$ . We do not yet fully understand this behavior, but we
are studying it both analytically and numerically and will
present our findings in a subsequent paper.

We will now do an extensive scan of parameter space $\{\mu_\phi,
\mu_\psi\}$ in order to have reliable estimates on the magnitude
and wavelength of the perturbations. This is important both to
make sure that CMB constraints can be satisfied as well as to
study the formation of primordial black holes that might lead to
the supermassive black holes found in the centers of galaxies.
Since the original motivation for this paper has been the
supersymmetric models first presented in \cite{sol1} and
\cite{sol2}, we will present the results for the perturbations in
these models. However, our quadratic approximation holds for more
general hybrid inflation models. Hence it is important to make a
model-independent parameter sweep. This will provide a more
general set of predictions of this class of models. We will give
both exact power spectra, as well as try to isolate the dominant
features and provide a qualitative understanding of their
dependence on the model's parameters. 

%%%%%%%%%%%%%%%%%%%%%%%%%%%%%%
%%%%%%%%%%%%%%%%%%%%%%%%%%%%%%
%%%%%%%%%%%%%%%%%%%%%%%%%%%%%%

\subsection{Model-Independent Parameter Sweep}

\begin{figure}
\centering
\begin{tabular}{cc}
\includegraphics[width=3in]{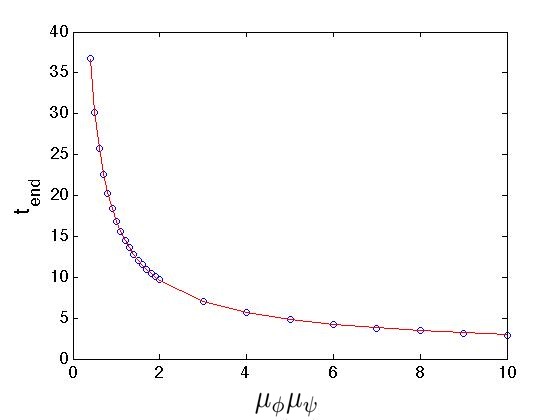}&
\includegraphics[width=3in]{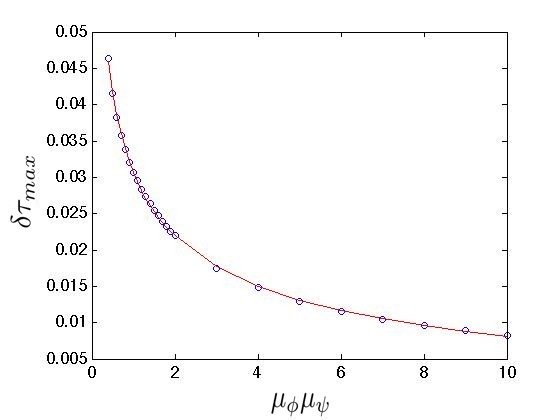} 
\\
\includegraphics[width=3in]{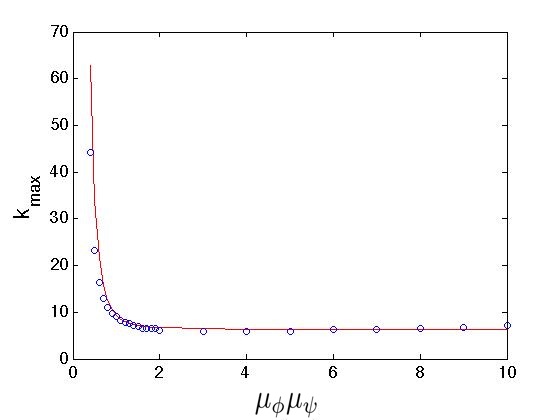}&
\includegraphics[width=3in]{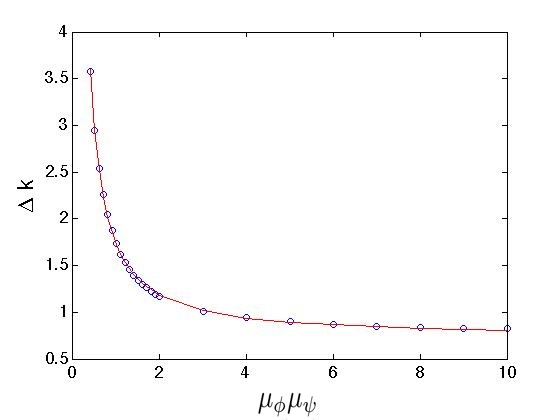} 
\end{tabular}
   \caption{Parameter sweep for constant timer field mass
$\mu_\psi =1/20$ and constant end field value $\phi_{\rm
end}=10^{14}$. Data points are plotted along with a least square
power law fit. The same trend is evident in all curves. The time
delay spectrum grows in amplitude and width and is shifted
towards larger momentum values as the mass product decreases.
Also inflation takes longer to end for low mass product.}
   \label{fig: param1}
\end{figure}

There are several model-dependent parameters that give us some
control over the properties of the resulting power spectrum.
Initially we will fix the value of the field at the end of
inflation to be $|\phi_{\rm end}| =10^{14}$ in units of the
Hubble parameter. With this assumption (which will be relaxed
later), we can calculate the properties of the power spectrum as
a function of the masses. Initially we fix the reduced timer
field mass to be $\mu_\psi = {1\over 20}$ and vary the mass of
the waterfall field. The results are shown in Fig.~\ref{fig:
param1}. We have plotted (clockwise from the top left)
\begin{enumerate}
\item The end time of inflation, defined as the time when the RMS
value of the field reaches the end value.
\item The maximum amplitude of the spectrum of the time delay.
\item The comoving wavenumber at which the aforementioned maximum
value occurs. Thinking about black holes, this is the scale at
which black holes will be most likely produced.
\item The width of the time delay distribution in the logarithmic
scale, taken as $\Delta k = \log_{10} \left ( {k_{+1/2} \over
k_{-1/2}} \right )$ where $k_{\pm 1/2}$ are the wavenumbers at
which the distribution reaches one half of its maximum value.
\end{enumerate}
We see that all the plotted quantities show a decreasing behavior
as one increases the mass product. In order to quantify this
statement, we fitted each set of data points with a power law
curve of the form $y=a x^b +c$. The scaling exponent $b$ for the
various quantities was $b_{t_{\rm end}} \approx -0.88$,
$b_{\delta\tau_{max}} \approx -0.34$, $b_k{_{max}} \approx
-3.219$, $b_{\Delta k} \approx -1.17$. As a comparison, the
corresponding best fit exponent of the growth rate $\lambda$ as a
function of the mass product is $b_\lambda \approx -0.85$.

\begin{figure}
\centering
\begin{tabular}{cc}
\includegraphics[width=3in]{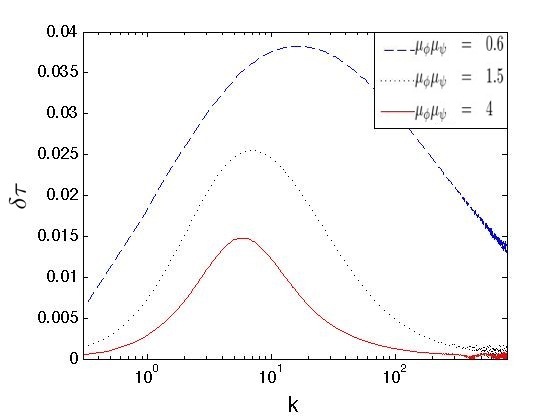}&
\includegraphics[width=3in]{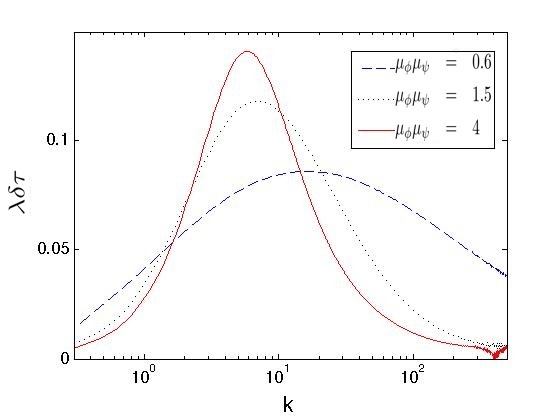} 
\end{tabular}
   \caption{Time delay spectra for different values of the mass product, keeping the timer field mass fixed at $\mu_\psi={1\over 20}$ }
   \label{fig:sample}
\end{figure}

In order to get a better understanding of what these parameters
actually mean, we plot three characteristic spectra for three
values of the mass ratio in Fig. \ref{fig:sample}. We also
rescale the spectra by the growth factor $\lambda$. This probes
the actual form of the two point correlation function, as seen in
momentum space. That is, it shows the evaluation of the spectrum
in Eq.~(\ref{eqn:powerspec}), while ignoring the factor of
$1/\lambda^2$ in the evaluation of $\left < \delta t(\vec x)
\delta t(\vec 0) \right >$ from Eq.~(\ref{eqn:delta-tau-2pt}).

\bigskip

\begin{figure}
\centering
\begin{tabular}{cc}
\includegraphics[width=3in]{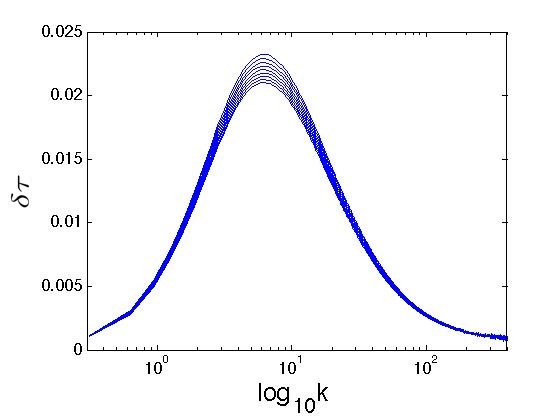}&
\includegraphics[width=3in]{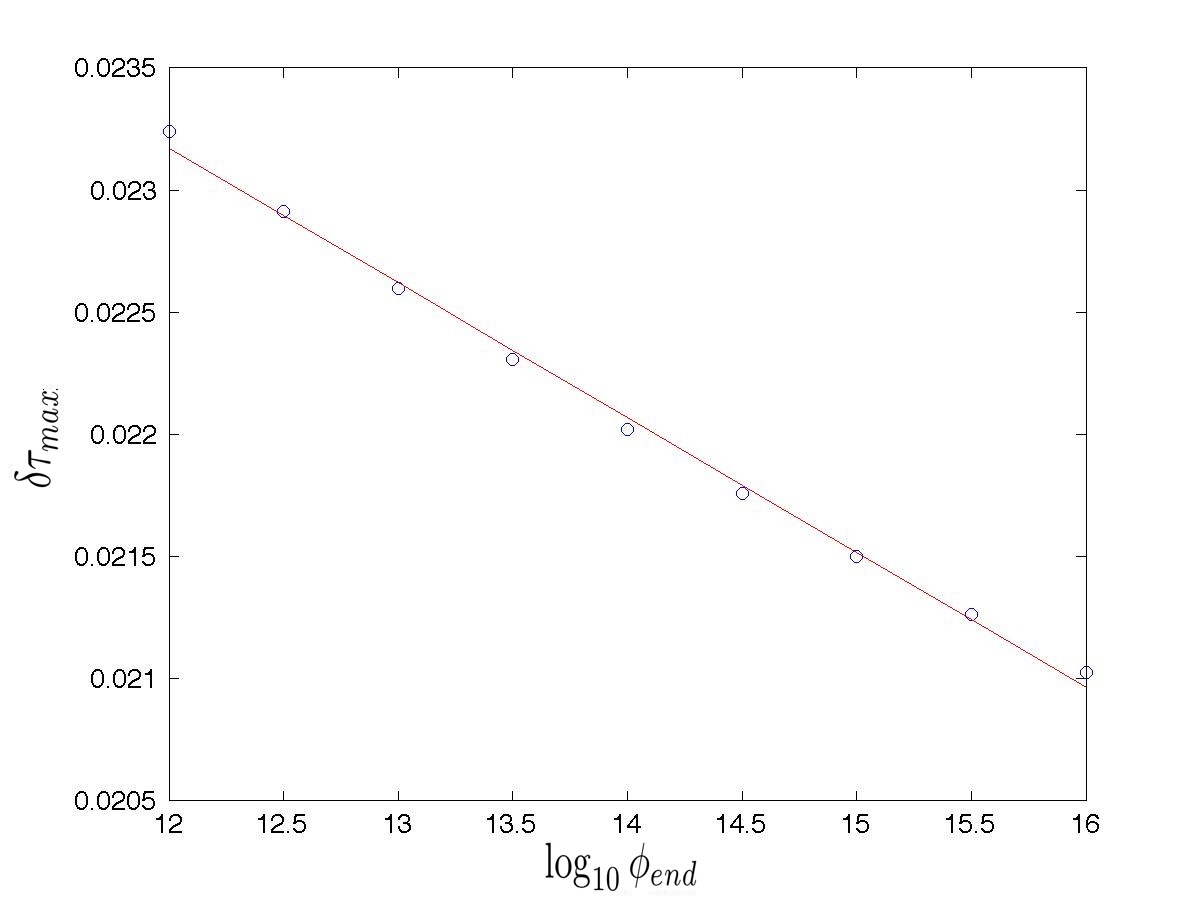} 
\\
\includegraphics[width=3in]{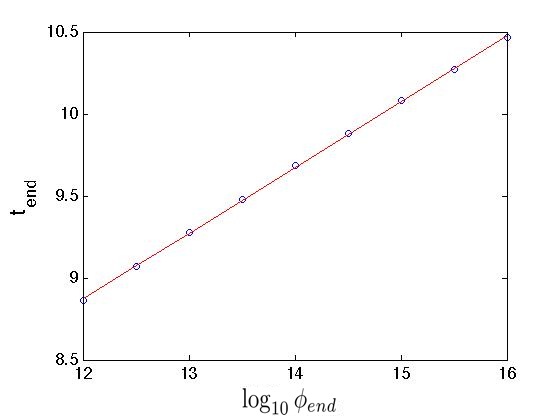}&
\includegraphics[width=3in]{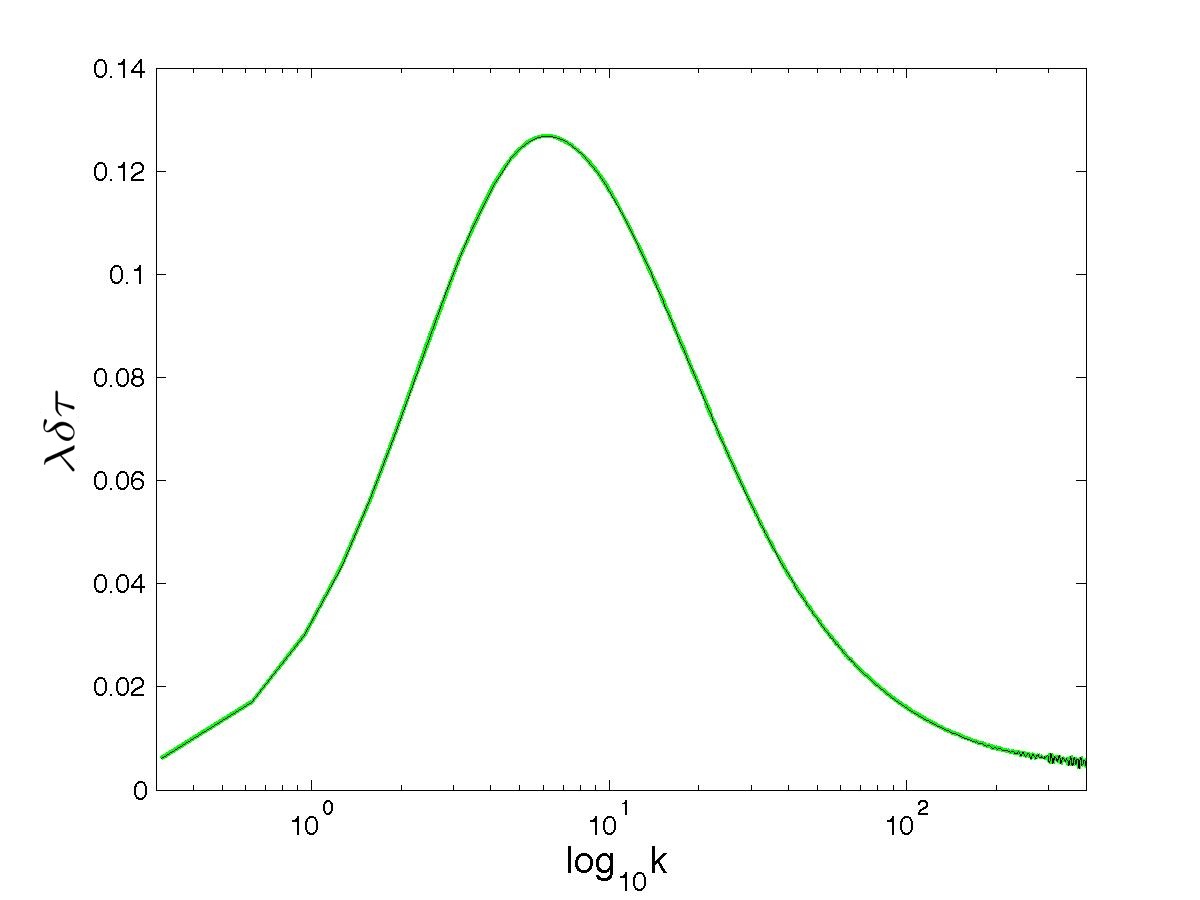} 
\end{tabular}
   \caption{Perturbation spectrum for varying field value at
$t_{\rm end}$ for constant masses. The time delay curves are
identical in shape and differ only in amplitude. This variation
is entirely due to the different value of the time dependent
growth factor $\lambda$, which differs for each case because
inflation simply takes longer to end for larger end field values.
}
   \label{fig: endfield}
\end{figure}

Before continuing to a more thorough examination of parameter
space, let us understand how changing the field value at the end
of inflation will change our results. Fixing the product of the
reduced masses equal to 2 ($\mu_\psi = {1\over 20}$ and
$\mu_\phi=40$), we let the field value $\phi_{\rm end}$ vary by
four orders of magnitude. The results are shown in Fig.
\ref{fig: endfield}. It is evident that the curves 
for $\delta \tau(k)$ are of identical form and slightly different
magnitude. The last graph shows the product $\lambda
\cdot \delta \tau (k) $ for the two curves at $\phi_{\rm end} =
10^{12}$ and $\phi_{\rm end} = 10^{16}$, plotted respectively as
a green thick and a black thin line. It is seen, that once
rescaled the two curves fall exactly on top of each other,
meaning that the actual integral that gives us the two point
function in position space is time independent, once we enter the
region where all modes behave identically. Furthermore if one
takes the product of the maximum value of the time delay times
the growth parameter ($\delta \tau_{max} \cdot
\lambda$) for the different values of 
$\phi_{\rm end}$ the result is constant for the range explored
here to better than 1 part in $10^6$, meaning that they are
identical within the margins of numerical error.  Thus, changing
the value of the field at which inflation ends can affect the
resulting perturbation spectrum only by changing the growth
parameter $\lambda$, for which we have a very accurate analytical
estimate in the form of Eq.~(\ref{lambda(N)}). From this point
onward, we will keep the end field value fixed at $\phi_{\rm
end}=10^{14}$ and keep in mind that the fluctuation magnitude can
change by $10\%$ or so if this field value changes. 

\bigskip

Once we fix the field magnitude at the end of inflation we have
two more parameters to vary, namely the two masses: the actual
timer field mass and the asymptotic tachyonic waterfall field
mass. The two masses can be varied either independently on a two
dimensional plane or along some line on the plane, in a specific
one-dimensional way. Fixing one of the two masses is such a way
of dimensional reduction of the available parameter space, as we
did before. Another way to eliminate one of the variables is to
fix the mass product and change the mass ratio. This will prove
and quantify the statement, that (at least for heavy waterfall
and light timer fields) the result is controlled primarily by the
mass product.

\begin{figure}
\centering
\begin{tabular}{cc}
\includegraphics[width=3in]{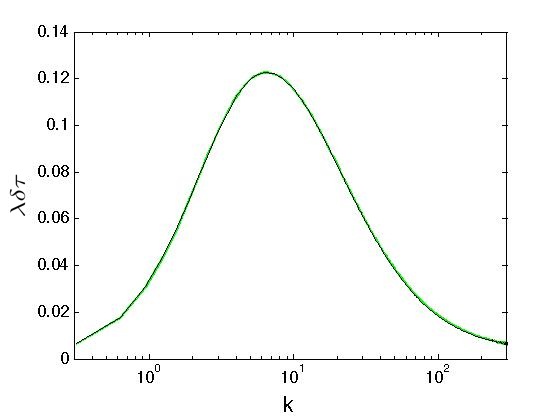} 
\includegraphics[width=3in]{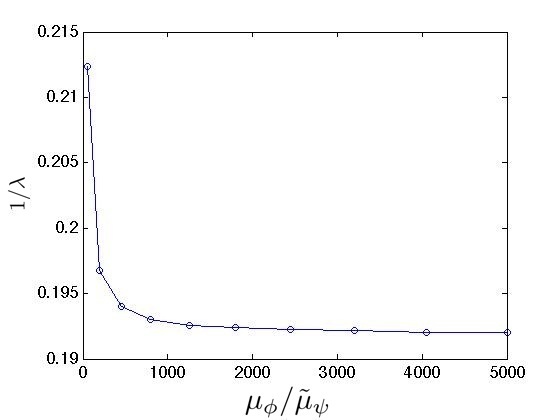}&
\\
\includegraphics[width=3in]{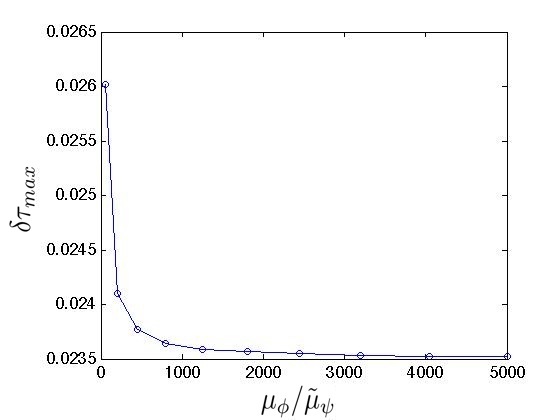} 

\includegraphics[width=3in]{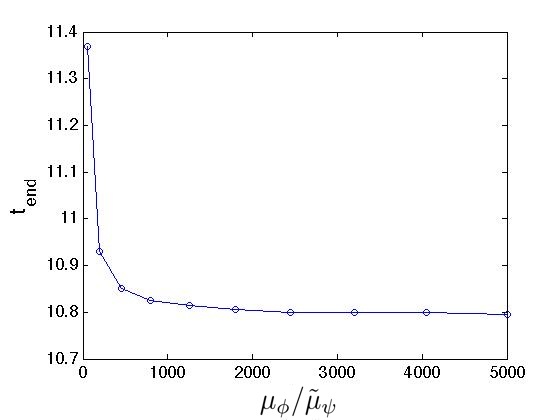}&
\end{tabular}
   \caption{Fixing the mass product at 2 and varying the mass
ratio. There is significant variation only for low mass ratio,
when the light timer field approximation loses its validity.
Furthermore the curves of maximum time delay amplitude and
$1/\lambda$ follow each other exactly up to our numerical
accuracy. Finally by rescaling the spectra by the growth factor
$\lambda$ they become identical for all values of the mass
ratio.}
   \label{fig: masspr2}
\end{figure}

We fix the mass product at $\mu_\phi\mu_\psi=2$. The results are
shown in Fig. \ref{fig: masspr2}. The curves are of identical
form and everything is again controlled only by $\lambda$. On the
top left figure we plotted $\lambda~\delta\tau$ for the two
extreme values and the curves fall identically on top of each
other (color-coding is as before). Furthermore if we calculate
the product $\lambda \cdot \delta\tau_{max}$ for different values
of the mass ratio we get a constant result $0.1225 \pm 2\cdot
10^{-5}$ where the discrepancy can be attributed to our finite
numerical accuracy.  The second feature of this calculation is
the extremely flat part of the end-time, growth factor and
maximum time delay curves for large values of the mass ratio and
the abrupt change as the mass ratio gets smaller. For the value
of the mass product that we have chosen, this transition happens
as the timer field mass approaches unity. Let us look at the
expansion of the effective waterfall field mass

\beq
\mu_{\phi,eff}^2 = \mu_\phi^2 \left ( 1-e^{-\tilde \mu_\psi^2 N} \right ) =  \mu_\phi^2 \tilde \mu_\psi^2 N (1-\tilde \mu_\psi^2 N + ...)
\eeq

When the second term in the expansion cannot be neglected, the
dynamics of the problem stops being defined by the mass product
alone. This explains the abrupt change we see as we lower the
mass ratio. By doing the same simulation for different values of
the fixed mass product we get similar results. 

\begin{figure}
\centering
\begin{tabular}{cc}
\includegraphics[width=3in]{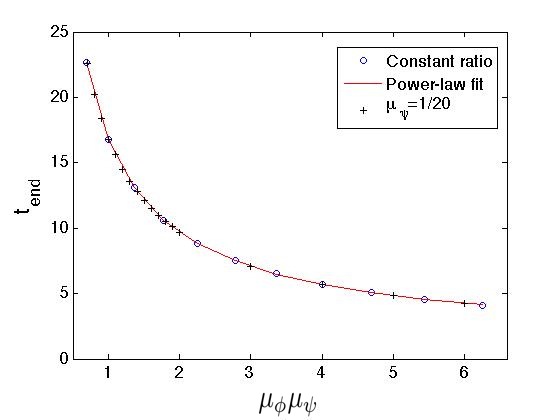}&
\includegraphics[width=3in]{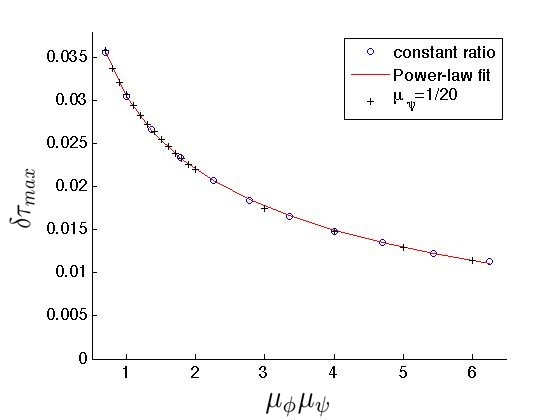} 
\\
\includegraphics[width=3in]{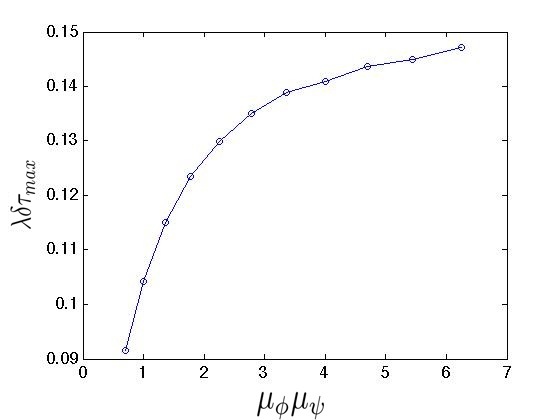}&
\includegraphics[width=3in]{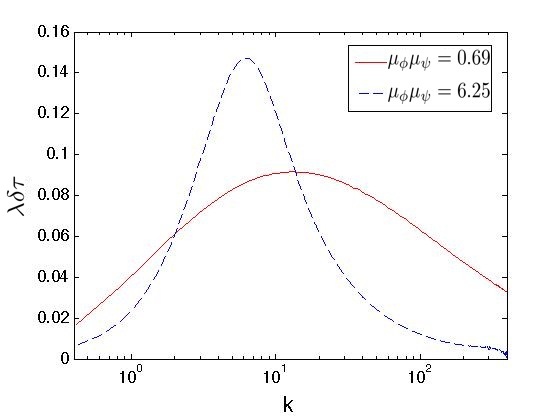} 
\end{tabular}
   \caption{ Fixing the mass ratio at 
900 (open circles) or the timer mass at $\mu_\psi = 1/20$ (+'s).
The time delay spectra for different mass products show no common
shape characteristics and remain different even when rescaled by
$\lambda$. Furthermore the end time and maximum perturbation
amplitude curves are identical for constant mass ratio and
constant timer field mass, proving that indeed the mass product
is the dominant parameter.}
   \label{fig: ratio}
\end{figure}

We can now do the opposite, that is fix the ratio and change the
mass product. The results are shown as the open circles in the
top two diagrams of Fig.~\ref{fig: ratio}, and in the lower
diagrams of the figure. There are two main comments to be made.
First of all, in the case of a fixed ratio, the growth rate
$\lambda$ does not solely determine the results. Rescaling the
spectrum by $\lambda$ not only fails to give a constant peak
amplitude (Fig.~\ref{fig: ratio}, lower left), but the result are
spectra of different shapes (Fig.~\ref{fig: ratio}, lower right).
On the other hand, the data points taken with a constant mass
ratio and a constant timer field mass ($\mu_\psi={1\over 20}$),
as shown by the +'s on the upper diagrams of Fig.~\ref{fig:
ratio}, fall precisely on the same curve! This clearly
demonstrates that the only relevant parameter, at least for a
light timer field, is the mass product!

We see that contrary to the fixed product case, the results for
fixed mass ratio do not depend solely on $\lambda$. We have
established that the the most important factor in determining the
time delay field is the product of the waterfall and timer field
masses, especially for a light timer field.

%%%%%%%%%%%%%%%%%%

\subsection{Supernatural Inflation}

We now turn our attention to the supernatural inflation models
that were studied in \cite {sol1} and \cite {sol2}. We will
examine each of the four cases separately.

\begin{figure}
\centering
\begin{tabular}{cc}
\includegraphics[width=3in]{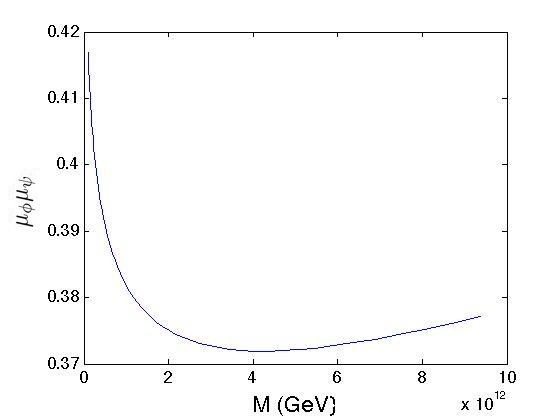}&
\includegraphics[width=3in]{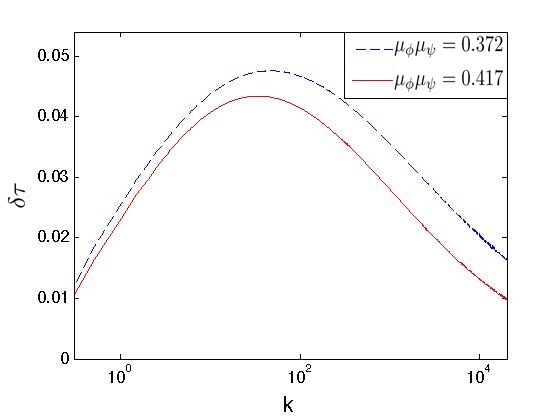} 
\end{tabular}
   \caption{First Supernatural inflation model with $M'$ at the
Planck scale. The spectra corresponding to the maximum and
minimum mass product are shown. We observe good agreement with
the results of the model independent parameter sweep of the
previous section, because the timer field mass is much smaller
than the Hubble scale.}
   \label{fig: susy1}
\end{figure}

Let us start with the first SUSY model (described by
Eq.~(\ref{SUSYModel1})) with the interaction-suppressing mass
scale $M'$ set at the Planck scale. The mass of the timer field
was calculated to be $50$ to $100$ times less than the Hubble
scale, while the asymptotic waterfall field mass was more than
$20$ times the Hubble scale. This means that the model is well
into the region where the two masses are separated by a few
orders of magnitude. According to the analysis of the previous
section, we expect the mass product to be the dominant factor in
the generation of density perturbations. In the left part of Fig.
\ref {fig: susy1} we see the mass product for this model. We can
see that the mass product varies less than $15\%$. It is hence
enough to calculate the time delay spectra for the two extreme
values and say that all other values of the mass product will
fall between the two, as shown in \ref{fig: susy1}.

Putting the mass scale $M'$ of the first SUSY model at the GUT
scale changes the masses as well as the Hubble scale by one order
of magnitude. However the reduced masses and their product have
very similar values as before. This is shown in Fig. \ref {fig:
susy2}

\begin{figure}
\centering
\begin{tabular}{cc}
\includegraphics[width=3in]{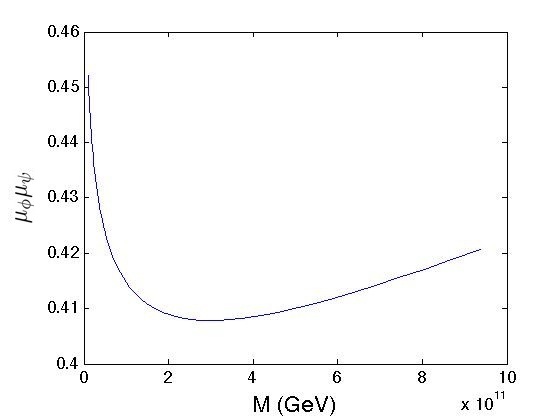}&
\includegraphics[width=3in]{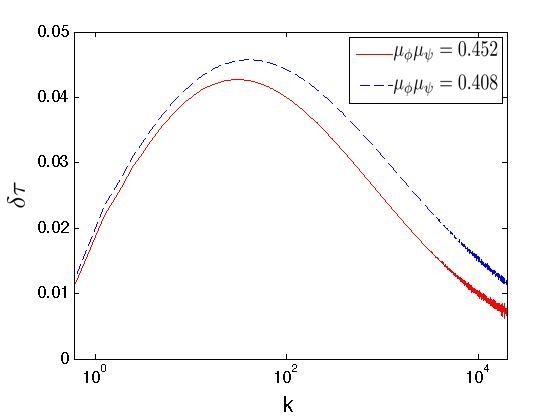} 
\end{tabular}
   \caption{First Supernatural inflation model with $M'$ at the
GUT scale. The spectra corresponding to the maximum and minimum
mass product are shown. There is again good agreement with the
results of the previous section.}  \label{fig: susy2}
\end{figure}

It is worth noting that these two SUSY models contain a very
light timer field, hence the results should be the same as our
previous parameter space sweep with a constant light timer field.
If one compares Fig. \ref{fig: susy1} and Fig. \ref{fig: susy2}
with Fig. \ref{fig: param1}, we indeed see excellent agreement
for the amplitude and width of the time delay spectrum. 

\begin{figure}
\centering
\begin{tabular}{cc}
\includegraphics[width=3in]{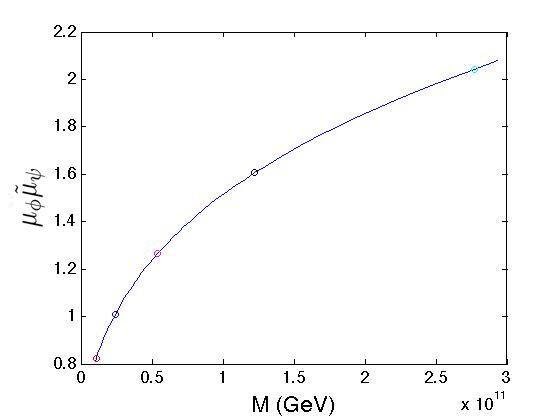}&
\includegraphics[width=3in]{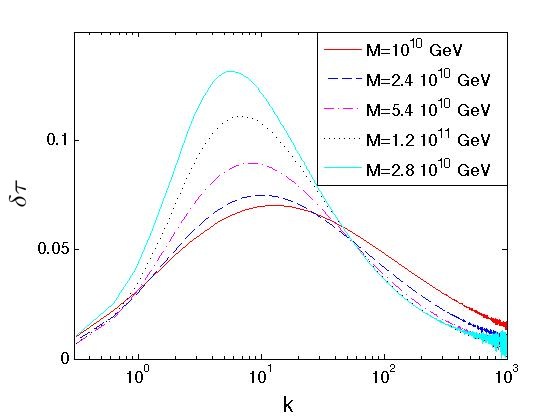} 
\end{tabular}
   \caption{First Supernatural inflation model with $M'$ at the
intermediate scale. Five representative pairs of masses were
chosen and the corresponding time delay curves are shown. This
model can give maximum time delay of more than $0.1$.}
   \label{fig: susy3}
\end{figure}

When setting the mass scale $M'$ at some lower scale of $10^{11}$
GeV, the reduced timer and waterfall masses become $O(1)$. This
means that in this case the parameter $\lambda$ saturates faster
and the perturbation spectrum reaches its asymptotic limit
earlier and becomes time-independent from that point onward.
Furthermore the actual value of the growth parameter $\lambda$ is
smaller, leading to an enhanced perturbation amplitude, the
largest among the models studied here. The mass product changes
by a factor of $2.5$ as seen in Fig.~\ref{fig: susy3}. We choose
five points in the allowed interval of mass values and calculate
the corresponding curves. The specific values of the mass
parameter $M$ are $M= 1.06 \cdot 10^{10} \hbox{ GeV},~ 2.4 \cdot
10^{10} \hbox{ GeV},~ 5.42 \cdot 10^{10} \hbox{ GeV},~ 1.23\cdot
10^{11} \hbox{ GeV},~ 2.77 \cdot 10^{11} \hbox{ GeV}$. The
corresponding pairs of reduced waterfall and timer masses are
$\{\mu_\phi,1/\mu_\psi \} = \{ 3.19 ,4.48\},~\{ 2.58 ,2.97\} ,~\{
2.22 ,2.05 \},~\{ 1.99, 1.47 \} ,~\{ 1.83, 1.09\}$. The points on
the mass product graph are color coded to match the corresponding
time delay curve in Fig. \ref {fig: susy3}. We can see that since
the mass products have a larger variation, the resulting spectra
have quite different time delay spectra. Also, since the timer is
not much lighter than the Hubble scale, the curves do not scale
according to our previous analysis. 

\begin{figure}
\centering
\begin{tabular}{cc}
\includegraphics[width=3in]{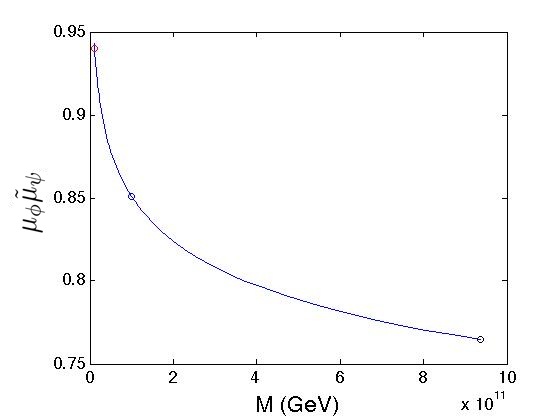}&
\includegraphics[width=3in]{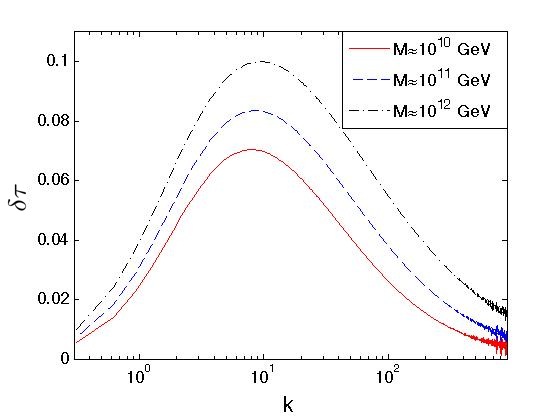} 
\end{tabular}
   \caption{Second Supernatural inflation model.  Three representative pairs of masses were chosen and the corresponding time delay curves are shown.}
   \label{fig: susyr2}
\end{figure}

We finally consider SUSY model 2, Eq.~(\ref{SUSYModel2}), with
the $\psi^2 \phi^2$ interaction term. Again the reduced masses
are $O(1)$, so we expect a small $\lambda$ leading to a large
amplitude perturbation spectrum. The mass product varies around
$1$ by less than $\pm 15\%$. We choose three values of the mass
product (the two extrema and an intermediate one) and plot the
resulting curves in Fig. \ref{fig: susyr2}.  The specific values
of the mass parameter $M$ are $M= 1.080 \cdot 10^{10} \hbox{
GeV},~ 1.006 \cdot 10^{11} \hbox{ GeV},~ 9.376 \cdot 10^{11}
\hbox{ GeV}$ and the corresponding pairs of reduced waterfall and
timer masses are $\{\mu_\phi,1/\mu_\psi \} = \{ 2.697, 2.367
\},~\{ 2.330 , 2.262\} ,~\{ 2.007, 2.170 \}$.

%%%%%%%%%%%%%%%%%%%%%%%%%%%%

%%%%%%%%%%%%%%%%%%%%%%%%%%%%
%%%%%%%%%%%%%%%%%%%%%%%%%%%%
%%%%%%%%%%%%%%%%%%%%%%%%%%%%
%%%%%%%%%%%%%%%%%%%%%%%%%%%%

\section{Conclusions}
We presented a novel method for calculating the power spectrum of
density fluctuations in hybrid inflation, one that does not
suffer from the non-existence of a classical field trajectory. We
used this method to numerically calculate the power spectrum for
a wide range of parameters and concluded that in the case of a
light timer field, all characteristics of the power spectrum are
controlled by the product of the masses of the two fields. In
particular the amplitude was fitted to a power law and found to
behave as $\delta\tau_{max} \sim 0.03 (\mu_\phi
\mu_\psi)^{-0.34}$ and the width in log-space as $\Delta k \sim
1.7 (\mu_\phi \mu_\psi)^{-1.17}$. Furthermore we made connection
to SUSY inspired models of hybrid inflation and gave numerical
results to their power spectra as well. For the SUSY models with
a light timer field the numerical results were in excellent
agreement with our fitted parameters.

Work is currently under way in refining and extending the
formalism. Understanding the rescaling properties between the RSG
approximation and the exact result could provide further insight
into the physics of the problem and provide quasi-analytical
approximation of well controlled accuracy. We will also apply our
results to estimating the number and size of primordial black
holes and try to make contact with astrophysical observations
regarding supermassive black holes in galactic centers. Finally
we are examining the predictions of our model for the
non-Gaussian part of the perturbation spectrum.

\section{ Acknowledgements }
We thank Kristin Burgess and Nguyen Thanh Son, whose earlier work
on this subject paved the way for the current project.  We also
thank Larry Guth, who helped us understand that this problem does
not require a Monte Carlo calculation, and Mark Hertzberg, who
helped us understand the dynamics of the waterfall transition. We
also thank Alexis Giguere, Illan Halpern, and Matthew Joss for
helpful discussions. The work was supported in part by the DOE
under Contract No. DE-FG02-05ER41360.

%%%%%%%%%%%%%%%%%%%%%%%%%%%%%%
%%%%%%%%%%%%%%%%%%%%%%%%%%%%%%
%%%%%%%%%%%%%%%%%%%%%%%%%%%%%%

\appendix

\section{Zero mode at early times}
\label{AppZero}

The $\vec k =0$ mode is not captured by the procedure described
in the main text. If we consider this mode alone for
asymptotically early times, so that we keep only the exponential
in the mass term
\beq
\ddot u +3\dot u = -\mu_\phi^2e^{-\tilde \mu_{\psi}^2N}u
\eeq
This can again be solved in terms of Bessel functions by defining
a new variable and a new function as
 \beq
 \tilde z = \alpha e^{-\tilde \mu_{\psi}^2N/2} ~~,~~   u(0,N)=\tilde z ^\beta \tilde Z(\tilde z)
\eeq
The mode function becomes
\beq
\tilde z ^2 {d^2 \tilde Z \over d \tilde z^2} 
+\tilde z {d \tilde Z \over d \tilde z} \left ( 1+2\beta -{6\over
\tilde \mu_\psi^2}\right ) + \tilde Z \left ( \beta^2 -{6\beta
\over \tilde \mu_\psi^2}+{\mu_\phi^2\tilde z^2 4\over
\alpha^2\tilde \mu_\psi^4}\right ) =0
\eeq
The standard form of the differential equation that gives Bessel
functions is
\beq
z^2 {d^2 Z_\nu \over dz^2} +z {dZ_\nu \over dz} +(z^2-\nu^2)
Z_\nu=0
\label{besse}
\eeq
By appropriately choosing the constants $\alpha$,$\beta$ and
$\nu$ the two equations can be made identical. The choices are
\beq
\beta = {3\over \tilde \mu_\psi^2} ~,~ \alpha = {2\mu_\phi \over \tilde \mu_\psi ^2}~,~\nu = {3\over \tilde \mu_\psi^2}
\eeq
Finally introducing an arbitrary constant of normalization $N_0$,
the solution for the zero mode at asymptotically early times
becomes
\beq
u(0,N) = N_0 e^{-3N/2} H_\nu ^{(1)} (\tilde z) ~~,~~ \tilde z =
{2\mu_\phi \over \tilde \mu_\psi ^2} e^{-\tilde \mu_{\psi}^2N/2}
\eeq
The normalization factor $N_0$ can be defined using the Wronskian
at early times. The Wronskian at all times is defined as
\beq
W(\vec k,t) = u(\vec k,t) {\partial u^*(-\vec k ,t) \over
\partial t} - {\partial u(\vec k ,t) \over \partial t} u^*(-\vec
k,t)
\eeq
Taking the time derivative and using the equation of motion
\beqn
\nonumber
{\partial W(\vec k ,t ) \over \partial t} &=& u(\vec k,t)
{\partial^2 u^*(-\vec k,t) \over \partial t^2}- {\partial^2
u^(\vec k,t) \over \partial t^2}u^*(-\vec k,t) = -3HW(\vec k,t)
\\
&& \Rightarrow W(\vec k,t) = f(\vec k) e^{-3Ht}
\eeqn
Since $f(\vec k) $ is by definition independent of time, we will
compute it at approximately early times, where we know the
solution in analytic form and the solution is
\beq 
W(\vec k \ne 0) = ie^{-3N} ~~,~~ W(\vec k =0) ={ 2ir\mu_{\psi}^2
H \over pi}N_0^2e^{-3N}
\eeq
Requiring that the Wronskian be a continuous function of $\vec k$
at all times we can extract the value of $N_0$.
 \beq
N_0 =\sqrt {3\pi \over 2r \mu_{\psi}^2 H}
 \eeq

\section{Initial Conditions}

We can rewrite the mode equation as a system of three coupled
first order differential equations.
\beqn
{d\theta \over dN} & = & -{\tilde k e^{-3N} \over R^2}
 \\
{dR \over dN} & = & \dot R
 \\
{d \dot R \over dN} & = & {\tilde k^2 e^{-6N} \over R^3} - 3\dot
R - e^{-2N} \tilde k^2 R +\mu_\phi ^2( 1-e^{-\tilde \mu_\psi
^2N})R
 \eeqn
In this notation, $\dot R$ is one of the three independent
functions. 

This is not a system of three coupled ODE's in the strict sense.
We can first solve the two equations $dR\over dN$ and $d\dot R
\over dN$ as they do not contain any terms involving $\theta$ or
its derivative. We can then integrate $d\theta \over dN$ forward
in time, using the calculated values of $R(N)$. Furthermore it is
clear that the equations only depend on the magnitude of the
wavenumber, as was expected due to the isotropy of the problem,
so we need only solve the mode equations for one positive semi
axis.

We know from the analytical solution at early times that
\beq
R(N\to-\infty) \to e^{-N}
\eeq
Since we have to start the numerical integration at some finite
negative time without losing much in terms of accuracy, we refine
the initial condition by including extra terms in the above
expression. We will then start numerically integrating when our
expansion violates the desired accuracy bound. We define the
correction to the asymptotic behavior as $\delta R(N)$ such that
\beq
R(N) \equiv e^{-N} + \delta R(N)
\eeq
We will expand $\delta R$ in powers of $\mu_\phi ^2$ and $e^{N}$
.
\begin{eqnarray}
\nonumber
\delta R && =\left ( {e^N\over 2\tilde k^2} -{e^{3N}\over 8\tilde
k^4} + {e^{5N}\over 16\tilde k^6} \right )+ \mu_\phi ^2 \left (
{e^N\over 4\tilde k^2} \left [ 1-e^{-\tilde \mu_\psi ^2 N} \right
] + {e^{3N}\over 16\tilde k^4} \left [ 4+ e^{-\tilde \mu_\psi ^2
N} (\tilde \mu_\psi^4 -6\tilde \mu_\psi^2 -4) \right ] \right .
 \\
 \nonumber
&& \left . -\mu_\phi ^2 {e^{5N}\over 64\tilde k^6} \left [ 86+
e^{-\tilde \mu_\psi ^2 N} (\tilde \mu_\psi^8 -14\tilde
\mu_\psi^6+53\tilde \mu_\psi^4-25 \tilde \mu_\psi^2 -86) \right ]
\right )
 \\
 \nonumber
&&+ 5 \mu_\phi ^4 \left( {e^{3N}\over 32\tilde k^4} \left [
1-e^{-\tilde \mu_\psi ^2 N} \right ]^2 - {e^{5N}\over 64\tilde
k^6} \left [ 29-e^{-\tilde \mu_\psi ^2 N}(9\tilde \mu_\psi^4-65
\tilde \mu \psi ^2 +58) \right. \right.
\\
&& \left. \left. + e^{-2\tilde \mu_\psi ^2 N} (14 \tilde
\mu_\psi^4-65 \tilde \mu \psi ^2 +29) \right ] \right ) + 15
\mu_\phi^6 {e^{5N}\over 128\tilde k^6} \left [ 1-e^{-\tilde
\mu_\psi ^2 N} \right ] ^3
\end{eqnarray}
We can now choose the initial expansion for $R(N)$ to calculate
the expansion for the phase $\theta (N)$. 

The asymptotic behavior, given by the standard definition of the
Hankel functions is
\beq
\theta (N\to -\infty ) = \tilde k e^{-N} -\pi \Leftrightarrow  \theta (N\to -\infty ) + \pi = \tilde k e^{-N}
\eeq
We define $\tilde \theta \equiv \theta (N) + \pi \Rightarrow
\dot{ \tilde \theta} = \dot \theta$, in order to keep track of
the constant phase factor without carrying it through the
perturbation expansion.

Defining the corrections to the early time behavior of the phase
as
\beq
\tilde \theta = e^{-N} [\tilde k + \delta \theta(N)]
\eeq
we can construct a similar expansion as the one for $\delta R$.
Although our formalism does not require knowledge of $\theta
(N)$, we included it for completeness.

\begin{figure}
\centering
\begin{tabular}{cc}
\includegraphics[width=3in]{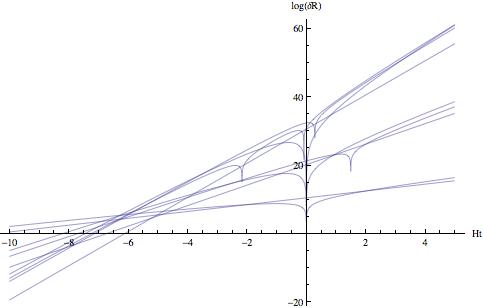}&
\includegraphics[width=3in]{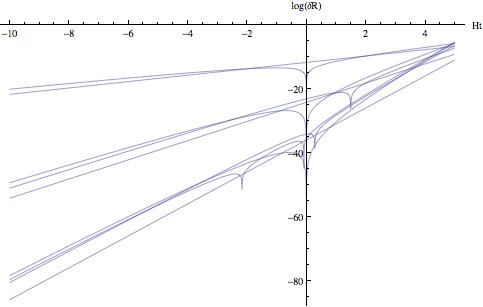} 

\end{tabular}
   \caption{Mode functions for $\mu_\phi=22$ and $\tilde
mu_\psi=1/18$. The left column is calculated for $\tilde k=1/256$
and the right for $\tilde k = 256$ }
   \label{fig: initial}
\end{figure}

%%%%%%%%%%%%%%%%%%%%%%%%%%%%%

\end{document}